\documentclass[prc,showpacs,floatfix]{revtex4}
\usepackage{bm}
\usepackage{dcolumn}
\usepackage{graphicx}
\usepackage{amsmath}
\usepackage{bbold}
\newcommand{\beq}{\begin{equation}}
\newcommand{\eeq}{\end{equation}}
\newcommand{\beqa}{\begin{eqnarray}}
\newcommand{\eeqa}{\end{eqnarray}}
\begin{document}
\title{
\hfill{\small {\bf MKPH-T-05-06}}\\
{\bf Incoherent pion photoproduction on the deuteron with
polarization observables II: Influence of final state rescattering}}
\author{A. Fix and H. Arenh\"ovel}
\affiliation{
Institut f\"ur Kernphysik,
Johannes Gutenberg-Universit\"at Mainz, D-55099 Mainz, Germany}
\date{\today}
\begin{abstract}
Incoherent pion photoproduction on the deuteron 
is studied for photon energies from threshold up to 1~GeV with special 
emphasis on polarization observables. The elementary 
$\gamma N\rightarrow\pi N$ amplitude is taken from the MAID model. 
The influence of final state interactions on total and semi-exclusive 
cross sections $\vec d(\vec\gamma,\pi)NN$ is investigated by including 
complete rescattering in the final $NN$- and $\pi N$-subsystems. For charged 
pion-production the influence of $NN$-rescattering is moderate whereas 
$\pi N$-rescattering is almost negligible. Much stronger influences of 
$NN$-rescattering are seen in neutral pion production, which is due to
the elimination of a significant spurious coherent contribution in the 
impulse approximation. Sizeable effects are also found in some of the 
beam, target and beam-target asymmetries of the differential cross section. 
\end{abstract}

\maketitle

\section{Introduction}
Photoproduction of pions on the deuteron has two main but complementary
points of interest. The first one is to obtain information on the
elementary reaction on the neutron by using the deuteron as an
effective neutron target. A prerequisite for this is that one has
reliable control on off-shell and medium effects. In order to minimize
such effects, quasi-free kinematics is preferred. The second but not 
secondary aspect is
just the influence of a nuclear environment on the production
process, for the study of which off-quasi-free kinematics is better
suited. 

Pion photoproduction on the deuteron has been studied quite extensively
over the past 50 
years, starting with early work in~\cite{ChL51,LaF52,BlL77}. The role
of final state interaction (FSI) has been investigated by
Laget~\cite{Lag78,Lag81} applying a diagrammatic approach. The
influence of FSI effects 
were found to be quite small for charged pion photoproduction compared
to the neutral channel. A satisfactory agreement with experimental
data was achieved for $\pi^-$ production~\cite{Be+73}. Subsequently, these
results were confirmed by Levchuk {\it et al.}~\cite{LeP96} for the 
$d(\gamma,\pi^0)np$ reaction for which the elementary photoproduction
operator of Blomqvist and Laget~\cite{BlL77} was used. This work was
improved and extended to charged pion production channels
in~\cite{LeS00}, where a more realistic elementary production 
operator from the SAID~\cite{Said} and MAID~\cite{Maid} multipole
analyses was taken and $NN$-rescattering included, based on the
Bonn r-space potential~\cite{MaH87}. The influence of $NN$-FSI was
confirmed and good agreement with experimental data was achieved. 
In the threshold region a sizeable effect from $\pi N$-rescattering
was noted in~\cite{Le+00} which arises from intermediate charged 
pion production with subsequent charge exchange rescattering on the
spectator nucleon. The influence of $NN$- and $\pi N$-rescattering 
on polarization observables has been investigated in~\cite{LeP96}
for the GDH sum rule in the $\pi^0np$-channel as well as in~\cite{LoS00}
for target asymmetries in the $\pi^-pp$-reaction. 

FSI effects in incoherent pion photoproduction were also
studied by Darwish et al.~\cite{DaA03a}. The same
approach was then applied to the spin asymmetry with 
respect to circularly polarized photons and vector polarized 
deuterons~\cite{DaA03b}, which determines the much discussed 
Gerasimov-Drell-Hearn sum rule~\cite{GDH}. 
However, the approach was limited to the
$\Delta(1232)$-resonance region in view of a relatively simple
elementary production operator, based on an effective Lagrangian
approach from Schmidt et al.~\cite{ScA96}. A puzzling result of this
work was that the influence of FSI on the total cross sections for
charged pion production resulted in a slight decrease in the
$\Delta$-resonance region in contrast to previous
work~\cite{Lag78,Lag81,LeS00} where a slight increase was
found. Recently, this work was extended in a series of
papers~\cite{Dar04a,Dar05a,Dar05b,Dar05c,DaS05} to a study of various
polarization asymmetries of the semi-exclusive differential cross
section for $\vec d(\vec \gamma,\pi)NN$. The semi-exclusive beam
asymmetry $\Sigma$ for linearly polarized photons and the target
asymmetries $T_{IM}$ with respect to polarized deuterons were considered
in~\cite{Dar04a} and beam-target asymmetries in~\cite{Dar05a}, in both
cases only in impulse approximation. Final state interaction effects
were subsequently discussed
in~\cite{Dar05b,Dar05c,DaS05}. Unfortunately, many of these results
are based on incorrect expressions for polarization observables as 
pointed out recently in~\cite{ArF05}, where a thorough derivation of
these observables is given.  

Thus the present work was motivated firstly to use a better elementary
production operator from the MAID model~\cite{Maid}, allowing one to go to
higher photon energies and also to give a more reliable description of
the threshold region. Secondly, we would like to clarify the role of 
the final state interaction (FSI) in view of the above mentioned 
differences in the role of FSI effects. Thirdly, the increasing
importance of polarization observables requires a more thorough and reliable
treatment as done in~\cite{Dar04a,Dar05a,Dar05b,Dar05c,DaS05}. In the
present work we consider again besides the impulse approximation (IA)
complete rescattering in the final two-body subsystems, i.e.\ in the
$NN$- and $\pi N$-subsystems. Results on the spin asymmetry of the
total cross section, based on the present approach, and its
contribution to the Gerasimov-Drell-Hearn sum rule have already been
reported in~\cite{ArF04}.  

In the next section we briefly review the basic formalism for the
general differential cross section with inclusion of polarization
observables as derived in~\cite{ArF05}. Furthermore,
the essential ingredients for the calculation of the $T$-matrix in the
impulse approximation (IA) and the rescattering contributions are
described here. The results on the unpolarized differential cross
section for the semi-exclusive process $\vec d(\vec \gamma,\pi)NN$ as
well as all beam, target, and beam-target asymmetries will be
presented and discussed in Sect.~\ref{results} together with a
comparison to existing data. Finally, we will conclude in
Sect.~\ref{outlook} with a summary and an outlook. The separation of the 
the various asymmetries of the semi-exclusive differential cross
section are discussed in Appendix~\ref{appa}, and
a modified impulse approximation is given in Appendix~\ref{appc}. 

\section{The formalism}
\begin{figure}[htb]
\includegraphics[scale=.7]{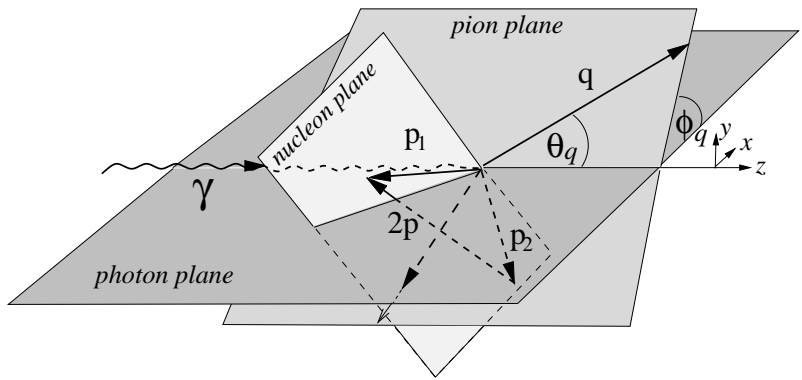}
\caption{Kinematics of pion photoproduction on the deuteron in the
laboratory system.}
\label{fig_kinematics}
\end{figure}

To begin with, we will briefly outline the kinematic framework of 
the reaction under study, namely
\beq
\gamma(k,\vec{\varepsilon}_\mu)+d(p_d)\!\rightarrow\!
\pi(q)+N_1(p_1)+N_2(p_{2})\,,
\eeq
defining the notation of the four-momenta of the participating
particles. The circular polarization of the photon is denoted by
$\vec{\varepsilon}_\mu$ ($\mu=\pm 1$). For the description of cross sections
and polarization observables we take as reference frame the 
laboratory frame and as independent variables for the characterization
of the final state
the outgoing pion momentum $\vec q=(q,\theta_q,\phi_q)$ and the
spherical angles $\Omega_p=(\theta_p,\phi_p)$ of the relative momentum
$\vec p=(\vec p_1-\vec p_2)/2=(p,\Omega_p)$ of the two outgoing
nucleons. The coordinate system is chosen as right-handed
with $z$-axis along the photon momentum $\vec{k}$. 
According to the convention of~\cite{ArF05}, we
distinguish in general three planes: (i) the photon plane spanned by
the photon momentum and the direction of maximal linear photon
polarization, which defines the direction of the $x$-axis, (ii) the
pion plane, spanned by photon and pion momenta, which intersects the
photon plane along the $z$-axis with an angle $\phi_q$, and (iii) the
nucleon plane spanned by total and relative momenta
of the two nucleons. It intersects the pion plane along the total
momentum of the two nucleons (see Fig.~\ref{fig_kinematics}). 
In case the linear photon 
polarization vanishes, one can choose $\phi_q=0$ and then photon 
and pion planes coincide. 

\subsection{The $T$-matrix}

All observables are determined by the $T$-matrix elements of the
electromagnetic pion production current $J_{\gamma\pi}$ between the
initial deuteron and the final $\pi NN$ states
\beq
T_{s m_s, \mu m_d}(q,\Omega_q,\Omega_p)=- ^{(-)}\langle \vec q,\vec p\,;s
m_s| \vec\varepsilon_\mu\cdot\vec J_{\gamma\pi}(0)|\vec d\,;1 m_d\rangle\,,
\eeq
where $s$ and $m_s$ denote the total spin and its projection on the
relative momentum of the outgoing two nucleons and $m_d$ correspondingly the
deuteron spin projection on the chosen $z$-axis. 
As is shown in~\cite{ArF05}, the dependence
on $\phi_q$ can be split of, i.e.\
\beq
T_{s m_s \mu m_d}(q,\Omega_q,\Omega_p)=e^{i(\mu+m_d-m_s)\phi_q} 
t_{s m_s \mu m_d}(q,\theta_q,\theta_p,\phi_{pq})\,.
\label{small_t}
\eeq
Thus the small $t$-matrix depends besides on $q$, $\theta_q$, and 
$\theta_p$ only on the difference of the azimuthal angles of 
$\vec q$ and $\vec p$, i.e.\ on $\phi_{pq}=\phi_{p}-\phi_{q}$. 
The small $t$-matrix elements are
the basic quantities which determine the differential cross section
and asymmetries. The latter are listed explicitly in~\cite{ArF05}.

For the calculation of the $T$-matrix we start from the impulse 
approximation (IA) to which
the contributions from $NN$- and $\pi N$-rescattering are
added. Possible two-body contributions to the electromagnetic
interaction are neglected. Thus the treatment is completely analogous
to previous work on incoherent $\pi$- and $\eta$-photoproduction on the
deuteron~\cite{DaA03a,FiA97} to which the reader is referred for formal
details. Then the $T$-matrix is given by the sum
\beq
T_{s m_s \mu m_d}=T_{s m_s \mu m_d}^{IA}+T_{s m_s \mu
m_d}^{NN}+T_{s m_s \mu m_d}^{\pi N}\,.
\eeq
For the IA contribution, which describes the production on one nucleon
while the other acts as a spectator, one has
\beqa
T_{sm_s\mu m_d}^{IA}&=&\langle\vec{q},\vec{p},\,sm_s\,|\,
\Big[t_{\gamma\pi}(1)+t_{\gamma\pi}(2)\Big]|\,1\,m_d\rangle
\nonumber\\
&=&\sum_{m_s^{\prime}}\Big(\langle sm_s\,|\,\langle \vec p_1|
t_{\gamma\pi}(W_{\gamma N_1})|-\vec p_2\rangle
\phi_{m_s^{\prime}m_d}(\vec{p}_2) 
|\,1\,m_s'\rangle-(1\leftrightarrow 2)\Big)
\,,
\eeqa
where $t_{\gamma\pi}$ denotes the elementary pion photoproduction
operator, which we take from the MAID model, $W_{\gamma N_1}$ the 
invariant energy of the $\gamma N$ system, and
$\vec p_{1/2}=(\vec k-\vec q\,)/2\pm\,\vec p$, and
$\phi_{m_sm_d}(\vec{p}\,)$ is related to the internal deuteron wave
function in momentum space by 
\beq
\langle \vec{p},
1m_s|1m_d\rangle^{(d)}=\phi_{m_sm_d}(\vec{p}\,)=\sum_{L=0,2}\sum_{m_L}i^L
(Lm_L\,1m_s|1m_d)u_L(p)\,Y_{Lm_L}(\hat{p})\,. \label{dwave}
\eeq
In view of the fact that for neutral pion production the major
influence of $NN$-rescattering arises from the non-orthogonality of the
final $NN$-plane wave to the deuteron ground state, we have considered
in addition a modified IA-amplitude, for which the deuteron component
is projected out from the final plane wave (see Appendix~\ref{appc}
for details). 

The two rescattering contributions have a similar structure
\beqa
T_{s m_s \mu m_d}^{NN}&=&\langle\vec{q},\vec{p},\,sm_s\,|\,
t_{NN}G_{NN}[t_{\gamma\pi}(W_{\gamma N_1})+t_{\gamma\pi}(W_{\gamma
N_2})]|\,1\,m_d\rangle\,,\\
T_{s m_s \mu m_d}^{\pi N}&=&\langle\vec{q},\vec{p},\,sm_s\,|\,
t_{\pi N}G_{\pi N}[t_{\gamma\pi}(W_{\gamma N_1})+t_{\gamma\pi}(W_{\gamma
N_2})]|\,1\,m_d\rangle\,,
\eeqa
where $t_{NN}$ and $t_{\pi N}$ denote respectively the $NN$ and $\pi
N$ scattering matrices and $G_{NN}$ and $G_{\pi N}$ the corresponding
free two-body propagators. For the actual evaluation, the scattering 
matrices are expanded into partial waves, and the expansion is then 
truncated at a certain angular momentum such that convergence is achieved.

\subsection{The differential cross section including polarization asymmetries}
\label{diff_cross}

The general five-fold differential cross section 
$d^5\sigma/dq d\Omega_q d\Omega_p$ including beam and 
target polarization has been derived in~\cite{ArF05}, and we refer
to this work for details. 
In the present work we are interested in the semi-exclusive reaction $\vec
d(\vec\gamma,\pi)NN$ where only the produced pion is detected. This means
integration of the five-fold differential cross section over $\Omega_p$,
yielding as semi-exclusive differential cross section~\cite{ArF05}
\beqa
\frac{d^3\sigma}{dq d\Omega_q}&=&
\frac{d^3\sigma_0}{dq d\Omega_q}
\Big[1+P^\gamma_l\,\Big\{\widetilde \Sigma^l\,\cos 2\phi_q
+\sum_{I=1}^{2} P^d_I \,\sum_{M= -I}^I 
\widetilde T_{IM}^l\cos[M\phi_{qd}-2\phi_q-\delta_{I1}\,\pi/2]
\,d^I_{M0}(\theta_d)\Big\}\nonumber\\&&
+\sum_{I=1}^{2} P^d_I \,\sum_{M= 0}^I
\Big(\widetilde T_{IM}^0\cos[M\phi_{qd}-\delta_{I1}\,\pi/2]
+P^\gamma_c\,\widetilde T_{IM}^c\sin[M\phi_{qd}+\delta_{I1}\,\pi/2]\Big)
\,d^I_{M0}(\theta_d)\Big]\,,\label{diffcrossc}
\eeqa
where $\phi_{qd}=\phi_q-\phi_d$. Explicit expressions for the asymmetries 
$\widetilde \Sigma^l$, $\widetilde
T_{IM}$, and $\widetilde T_{IM}^{c/l}$ are listed in the appendix 
of~\cite{ArF05}. Furthermore, the photon
polarization is characterized by the degree of circular polarization
$P^\gamma_c$ and the degree of linear polarization $P^\gamma_l$, where
the $x$-axis has been chosen in the direction of maximum linear
polarization. The deuteron target is characterized by four parameters,
namely the vector and tensor polarization parameters $P_1^d$ and
$P_2^d$, respectively, and by the orientation angles $\theta_d$ and
$\phi_d$ of the deuteron orientation axis with respect to which the
deuteron density matrix has been assumed to be diagonal. 

We would like to point out that in forward and backward pion
emission the following asymmetries vanish at $\theta_q=0$ or $\pi$
\beq
\widetilde \Sigma^l=0,\quad \widetilde T_{IM}^{0,c}=0\,\,\mbox{for}\,\,
M\neq 0,\quad\mbox{and}\quad  T_{IM}^{l}=0\,\,\mbox{for}\,\,M\neq 2\,,
\label{asym0}
\eeq
because of helicity conservation, i.e.\ in this case the cross section 
should not depend on $\phi_q$.

In the next section we will present results for the case that only the
direction of the outgoing pion is measured and not its momentum. Then
the corresponding differential cross section 
$d^2\sigma/d\Omega_q$ is given by an expression formally analogous to 
(\ref{diffcrossc}), where unpolarized cross section and asymmetries are 
replaced by
\beqa
\frac{d^3\sigma_0}{dq d\Omega_q}\,&\rightarrow&\frac{d^2\sigma_0}{d\Omega_q}=
\int_{q_{min}(\theta_q)}^{q_{max}(\theta_q)} dq\,
\frac{d^3\sigma_0}{dq d\Omega_q}\,,\\
\frac{d^3\sigma_0}{dq d\Omega_q}\,\widetilde \Sigma^l(q,\, \theta_q)
&\rightarrow&\frac{d^2\sigma_0}{d\Omega_q}\,\Sigma^l(\theta_q)=
\int_{q_{min}(\theta_q)}^{q_{max}(\theta_q)} dq\,
\frac{d^3\sigma_0}{dq d\Omega_q}\,\widetilde \Sigma^l(q,\, \theta_q)\,,\\
\frac{d^3\sigma_0}{dq d\Omega_q}\,\widetilde T_{IM}^\alpha(q,\, \theta_q)
&\rightarrow&\frac{d^2\sigma_0}{d\Omega_q}\,T_{IM}^\alpha(\theta_q)=
\int_{q_{min}(\theta_q)}^{q_{max}(\theta_q)} dq\,
\frac{d^3\sigma_0}{dq d\Omega_q}\,
\widetilde T_{IM}^\alpha(q,\, \theta_q)\,,\quad \alpha\in\{0,l,c\}\,.
\eeqa
The upper and lower integration limits are given by
\beqa
q_{max}(\theta_q)&=& \frac{1}{2b}\,\Big(a\,\omega \cos \theta_q
+ E_{\gamma d}\sqrt{a^2-4b\,m_\pi^2}\Big)\,,\\
q_{min}(\theta_q)&=& \max\{0,\frac{1}{2b}\,\Big(a\,\omega \cos \theta_q
- E_{\gamma d}\sqrt{a^2-4b\,m_\pi^2}\Big)\}\,,
\eeqa
where
\beqa
a&=&W_{\gamma d}^2+m_\pi^2-4\,m_{N}^2\,,\\
b&=&W_{\gamma d}^2+\omega^2\sin^2\theta_q\,,\\
W_{\gamma d}^2&=&m_d(m_d+2\,\omega )\,,\\
E_{\gamma d}&=&m_d+\omega \,.
\eeqa
Finally, in the total cross section only a few polarization observables 
survive, namely one has~\cite{ArF05} 
\beq
\sigma(P^\gamma_l,P^\gamma_c,P^d_1,P^d_2)
= \sigma_0\,\Big[1+P^d_2\,\overline T_{20}^{\,0}\,\frac{1}{2}
(3\cos^2\theta_d-1)
+P^\gamma_c\,P^d_1\,\overline T_{10}^{\,c}\,\cos\theta_d
+P^\gamma_l\,P^d_2 \,\overline T_{22}^{\,l}\cos(2\phi_d)
\,\frac{\sqrt{6}}{4}\sin^2\theta_d\Big]\,,
\eeq
where unpolarized total cross section and asymmetries are given by
\beqa
\sigma_0&=&\int d\Omega_q \int_{q_{min}(\theta_q)}^{q_{max}(\theta_q)}
dq\,\frac{d^3\sigma_0}{dq d\Omega_q}\,,\\
\sigma_0\,\overline T_{IM}^{\,\alpha}&=&
\int d\Omega_q \int_{q_{min}(\theta_q)}^{q_{max}(\theta_q)}
dq\,\frac{d^3\sigma_0}{dq d\Omega_q}\,\widetilde T_{IM}^{\,\alpha}\,,
\eeqa
with $\alpha\in\{0,l,c\}$.
This concludes the formal part.

\section{Results and discussion}\label{results}

\begin{figure}[htb]
\includegraphics[scale=.9]{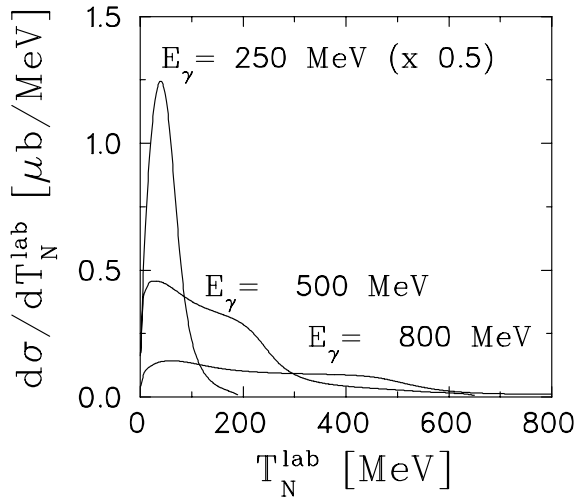}
\caption{Semi-exclusive differential cross section $d\sigma/dT_{N}$ 
in IA with respect to the equivalent nucleon laboratory kinetic energy 
$T_{N}$ of $NN$-scattering for three photon energies $E_\gamma=250$, 
500 and 800~MeV.}  
\label{fig_tnn_vertlg}
\end{figure}

For the calculation of the $NN$-rescattering contribution we have
taken the separable representation of the realistic Paris potential
from~\cite{HaP85} and included all partial waves up to 
$^3D_3$. Also the deuteron wave function was calculated using this
potential. In principle, any other realistic potential, e.g.\ the Bonn
r-space potential~\cite{MaH87}, could be used
as well, because the results do not depend sensitively on the
potential model as was found in~\cite{DaA03a}. However, one comment is
in order with respect to the question, whether the use of such a
nonrelativistic $NN$-potential can be justified in view of the high
energies involved, because the potential is fit to $NN$-scattering 
data up to nucleon lab kinetic energies of $T_{NN}=330$~MeV only. 
But even up to $T_{NN}=500$~MeV, it reproduces
the phase shifts reasonably well because the inelasticity parameters are
still small. Since the calculation of $NN$-rescattering requires  
$NN$-scattering amplitudes also at considerably higher energies, where 
we still use the same separable representation, one might expect some 
serious error. On the other hand, the size of this error depends on the 
relative size of that part of the phase space, where the energy
of the $NN$-subsystem exceeds the region of validity. In order to 
estimate the corresponding error, we present in Fig.~\ref{fig_tnn_vertlg} 
the cross section $d\sigma/dT_{N}$ as a function of the equivalent nucleon 
kinetic lab energy $T_{N}$ of $NN$-scattering. 
One readily notes that even for a photon energy of 800~MeV
the dominant part of the cross section corresponds to nucleon
laboratory kinetic energies of less than 500~MeV. Thus, the use of such a 
realistic potential is justified.

Similarly, $\pi N$-rescattering is evaluated using a
realistic separable representation of the $\pi N$-interaction 
from~\cite{NoB90} and taking into account all partial waves up to
$l=2$. 
We have evaluated the semi-exclusive differential cross section
including the various polarization asymmetries in IA alone and with
inclusion of $NN$- and $\pi N$-rescattering. As already mentioned, the
elementary pion photoproduction amplitude is taken from the MAID
model. Since it is parametrized in terms of the
Chew-Goldberger-Low-Nambu amplitudes (CGLN)~\cite{ChG57}, defined in
the $\gamma N$ c.m.\ frame, we first had to transform this amplitude
into a general frame of reference. This is achieved by introducing
invariant amplitudes and establishing relations to the CGLN amplitudes
in the c.m.\ frame. This is described in detail in~\cite{SaA04} and
needs not to be repeated here. 
Furthermore, for the evaluation of the MAID amplitudes the invariant 
$\pi N$-energy and the pion angle in the $\pi N$ c.m.\ system have to be
specified. For this purpose we assume that the four-momenta $q$ and $p_f$ of
pion and active nucleon in the final state obey the on-shell
condition. Then the corresponding c.m.\ variables are obtained by a Lorentz 
transformation with $\vec{\beta}=-(\vec{q}+\vec{p}_f)/(w_\pi+E_f)$. 
The initial four-momentum $p_i$ of the active nucleon is determined by 
assuming four-momentum conservation at the elementary vertex, i.e.\  
$p_i=q+p_f-k$.

\subsection{Total cross sections}
\begin{figure}[htb]
\includegraphics[scale=.9]{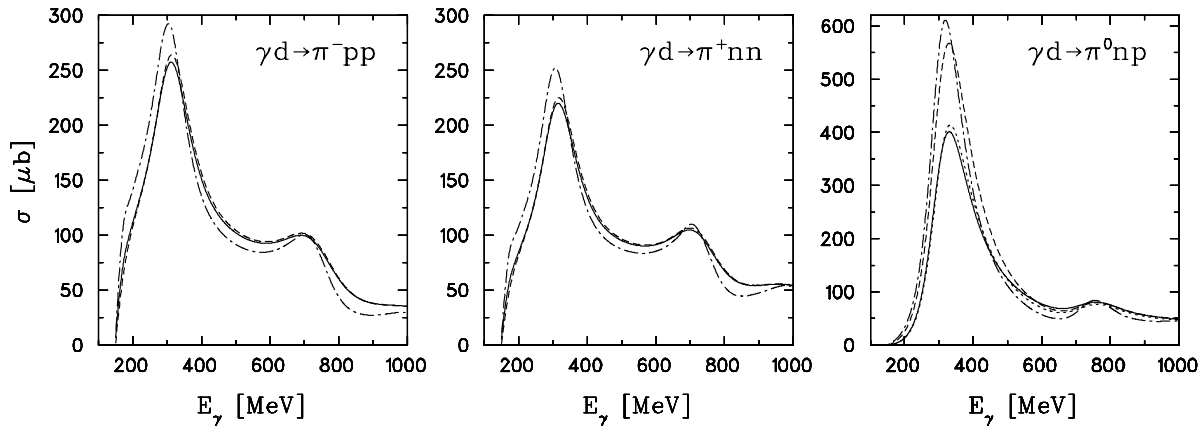}
\caption{Total cross section for pion photoproduction on the
deuteron. Notation of curves: dashed: IA; solid: IA + \protect$NN$-
and \protect$\pi N$-rescattering; dash-dot: corresponding elementary
cross section. For $\pi^0$-production: dotted curve describes the
modified IA.}  
\label{fig_tot}
\end{figure}

\begin{figure}[htb]
\includegraphics[scale=.7]{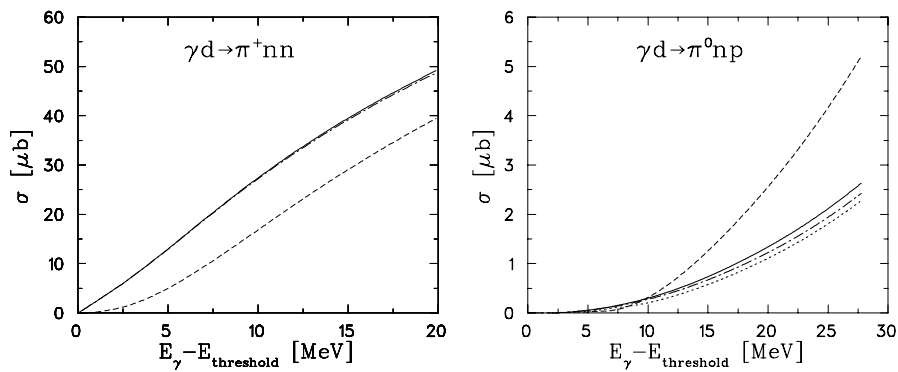}
\caption{Total cross section near threshold for \protect$\pi^+$- and 
\protect$\pi^0$-photoproduction on the deuteron. Notation of curves:
dashed: IA; dash-dot: IA + \protect$NN$-rescattering; 
solid: IA + \protect$NN$- and $\pi N$-rescattering; dotted in the 
right panel: modified IA.} 
\label{fig_tot_threshold}
\end{figure}

We begin the discussion of the results with the total cross sections 
for the three charge states displayed in Fig.~\ref{fig_tot}, where we 
have plotted in addition the corresponding elementary cross section for 
comparison. The threshold region is separately plotted in
Fig.~\ref{fig_tot_threshold}. 
One readily notes that for charged pion
production the rescattering effects are in general quite small. Only
close to threshold they lead to a sizeable enhancement (see left panel
of Fig.~\ref{fig_tot_threshold}), mainly by
$NN$-rescattering while $\pi N$-rescattering is almost negligible. 

The significant role of $NN$-FSI in photoproduction of mesons at very
low energies has been noted previously in~\cite{Nob67,Lag81,FiA97}. We
would only like to mention that this effect has a kinematical rather
than a dynamical origin. As pointed out in~\cite{FiA97}, 
in IA the energy needed for pion production below the free nucleon 
threshold is provided exclusively through the high momentum of a 
nucleon moving towards the incoming photon. As a result, the IA
predicts an anomalous suppression of the cross section, because of a
small probability for finding a nucleon with high momentum in the
deuteron wave function. Thus, $NN$-rescattering provides a mechanism
to balance the strong mismatch between the momentum needed to produce
the pion and the characteristic internal nucleon momentum in the
deuteron, so that the strong suppression appearing in IA can be
avoided. The same reasoning is true for $\pi^0$ channel. But in this
case the below mentioned role of orthogonality turns out to be very
important so that the resulting FSI effect becomes destructive already
at 10 MeV and higher above threshold. 

At higher energies, near and above the maximum the cross sections of 
charged pion production are reduced by FSI by a few percent. Therefore, 
the main difference 
to the elementary cross section comes from the Fermi motion leading to 
a slight reduction and a shift of the maximum, and a broadening of the
whole distribution. As one can see in Fig.~\ref{fig_tot}, the result
of these features is that the energy integrated cross section 
\begin{equation}
I(E)=\int\limits_{E_{th}}^E\sigma(E_\gamma)dE_\gamma
\end{equation}
as function of the upper integration limit $E$, is preserved over a wide 
energy region. In other words, the integral
at $E=1$~GeV has approximately the same value for the reaction on the free 
nucleon and on the deuteron. Concerning the role of FSI in the
$\pi^{\pm}$ channels at higher energies, one can assume that the
interaction between the emitted particles leads basically to a
redistribution of events in phase space, so that in the absence of 
absorption the overall yield of particles does not change and the FSI effect
in the total cross section remains insignificant. It is, however, not the case
for the near threshold region where, as was discussed above, the IA amplitude 
turns out to be anomalously suppressed in the available phase space.

\begin{figure}[htb]
\includegraphics[scale=.8]{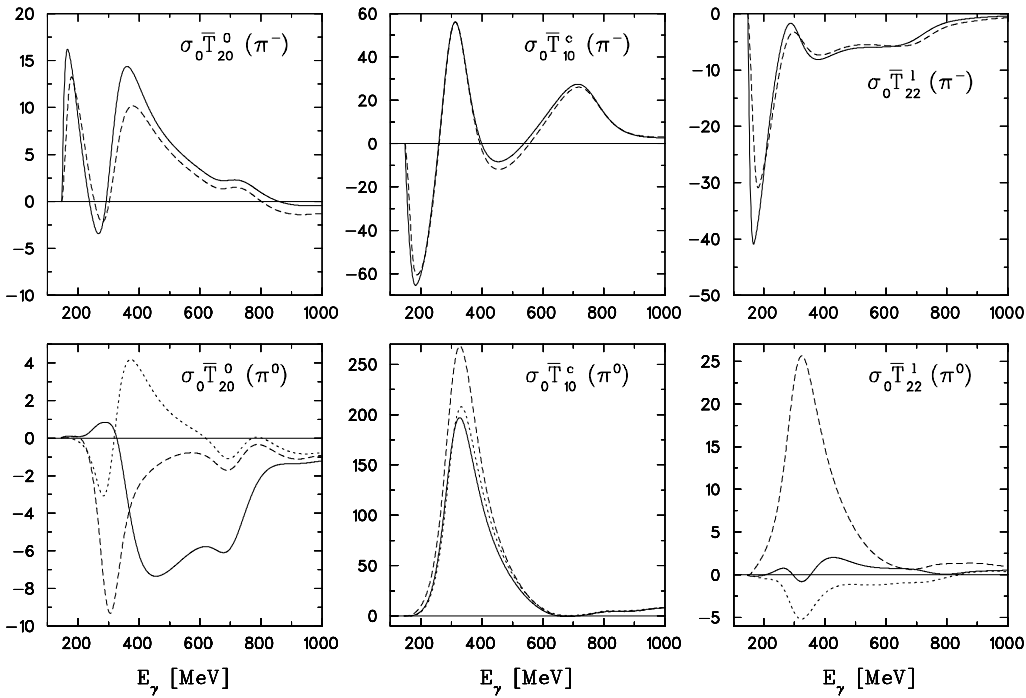}
\caption{Polarization asymmetries $\sigma_0\,\overline T_{20}^{\,0}$, 
$\sigma_0\,\overline T_{10}^{\,c}$, and $\sigma_0\,\overline T_{22}^{\,l}$ 
of the total cross section for $\pi^-$- (upper panels) and 
$\pi^0$-photoproduction (lower panels) on the
deuteron. Notation of curves: dashed: IA; solid: IA + 
\protect$NN$-rescattering; For $\pi^0$-production: dotted curve describes the
modified IA.}  
\label{fig_tot_asy}
\end{figure}
In contrast to charged pion production, one notes
quite large FSI effects from $NN$-rescattering in $\pi^0$-production as
displayed in the right panel of Fig.~\ref{fig_tot}. However, these 
effects are mainly due to the elimination of the spurious coherent
contribution in the IA. That is nicely demonstrated by the results for
the modified IA, also shown in Fig.~\ref{fig_tot}, where the deuteron 
wave function component in the final plane wave is projected out 
(see Appendix~\ref{appc} for details). The additional
rescattering contributions beyond the orthogonality effect are
comparable to what was found in charged pion production, except near
threshold where one finds a much smaller influence of FSI. With respect
to the elementary cross section, one notes a sizeable reduction for
incoherent $\pi^0$-production on the deuteron, because 
part of the strength goes to the coherent channel $d(\gamma,\pi^0)d$.
The spurious admixture of the coherent channel in the IA cross section 
was also discussed in~\cite{Sio01}.

The polarization observables of the total cross section are shown in 
Fig.~\ref{fig_tot_asy} only for $\pi^-$- and $\pi^0$-production since the 
ones for $\pi^+$- and $\pi^-$-production are quite similar. The vector 
asymmetry for circularly polarized photons 
$\sigma_0\,\overline T_{10}^{\,c}$ (middle panels), which governs the spin
asymmetry of the GDH sum rule, reproduces the results reported in~\cite{ArF04}.
The sensitivity to FSI is weak, similar to the total cross sections. The
tensor asymmetries for $\pi^-$-production are sizeable in the near threshold 
region, exhibiting a sharp peak in absolute values. 
$\sigma_0\,\overline T_{20}^{\,0}$ reaches a second quite broad maximum
above the $\Delta$-resonance, while $\sigma_0\,\overline T_{22}^{\,l}$ 
remains small. FSI is more notable than in the total cross section. 
For $\pi^0$ production, the tensor asymmetries show quite a different behavior
compared to $\pi^-$-production. They are quite small in general, almost
vanishing in the near threshold region. Furthermore, they show huge
FSI effects beyond the orthogonality effect. 

\subsection{Differential cross section}

\begin{figure}[htb]
\includegraphics[scale=.5]{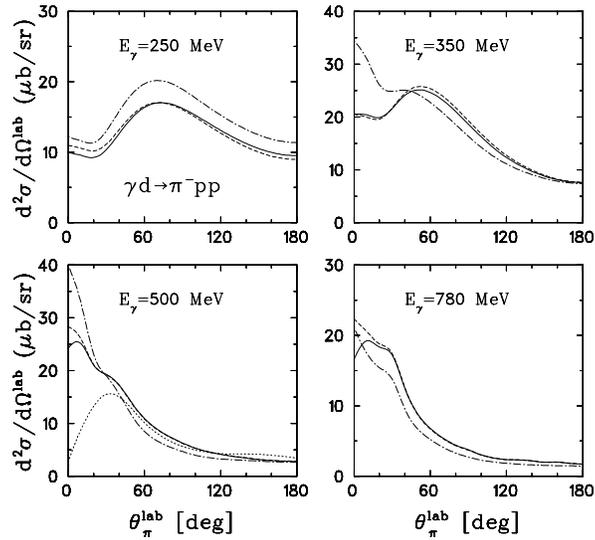}
\caption{Semi-exclusive differential cross section for \protect$\pi^-$
photoproduction on the deuteron. Notation of curves: dashed: IA;
solid: IA + \protect$NN$-rescattering; dash-dot: elementary process 
\protect$n(\gamma,\pi^-)p$; dotted in lower left panel: 
IA from~\protect\cite{LeS00}.} 
\label{fig_piMi_diff}
\end{figure}
\begin{figure}[htb]
\includegraphics[scale=.5]{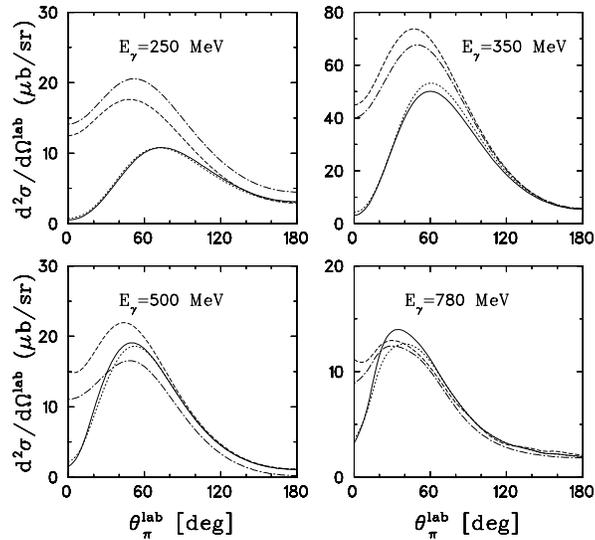}
\caption{Semi-exclusive differential cross section for \protect$\pi^0$
photoproduction on the deuteron. Notation of curves: dashed: IA;
dotted: modified IA; solid: IA + \protect$NN$-rescattering; dash-dot:
average elementary process on neutron and proton.}
\label{fig_pi0d_diff}
\end{figure}

Angular distributions are shown in Fig.~\ref{fig_piMi_diff} for
$\pi^-$-production and in Fig.~\ref{fig_pi0d_diff} for
$\pi^0$-production. Like for the total cross section, one notes for
$\pi^-$-production a small size of rescattering effects. At the lowest
energy (250 MeV) they 
decrease the cross section at forward angles and increase it in
backward direction. At 350 MeV one finds a slight enhancement near the
maximum, while at the two higher energies a sizeable decrease is found
only in the extreme forward direction. For comparison, we show also
the elementary cross sections, and again the smoothing effect of the
Fermi motion is apparent. 

Compared our calculation with the predictions of~\cite{Lag81} and 
\cite{LeS00} we note a significant disagreement for the IA at forward angles. 
At $E_\gamma$=500 MeV our IA cross section exhibits a visible 
rise which is governed mainly by the pion photoelectric 
term in the elementary amplitude. At the same time the calculations
of~\cite{LeS00} (see dotted curve in the lower left panel of 
Fig.~\ref{fig_piMi_diff} at
$E_\gamma$=500 MeV) predict a strong reduction in the same angular
region, so that the corresponding IA result at $\theta_\pi= 0$ is
about 10 times smaller than ours. 

A possible reason of this discrepancy has been discussed 
in~\cite{DaA03a}. Namely, it was claimed in~\cite{LeS00} that the drastic
reduction of the cross section compared to the elementary reaction is
a manifestation of the Pauli principle, leading to a strong
suppression of the $pp$ states in the region of small relative momenta
in this subsystem. On the contrary, we would like to note that this
suppression takes place only in the triplet state ($s=1$) of the
emitted protons, whereas the singlet part ($s=0$), where the processes
on the individual nucleons can coherently enhance each other, peaks at
$\theta_\pi=0$ similar to the elementary cross section. The resulting
angular distribution shows some reduction at forward angles, compared
to the elementary reaction, which, however, is not as large as that
exhibited by the triplet part only and is, therefore, much smaller
compared to the prediction of~\cite{Lag81,LeS00}.  

On the other hand, inclusion of FSI leads in~\cite{Lag81,LeS00} to a
significant increase at small angles compared to the IA cross section,
whereas in our case the effect of FSI is small and destructive. It is,
therefore, interesting to note that the difference between the full
results of the present work (see Fig.~\ref{fig_pimi_sig} below) and
of~\cite{LeS00} is rather small, so that both models describe the
data equally well. 

As was mentioned above, 
in the $\pi^0$ channel the orthogonality effect appears so that
the role of FSI is approximately determined by the fraction of the coherent 
reaction $\gamma d\to\pi^0 d$ in the inclusive $\pi^0$ photoproduction. 
Fig.~\ref{fig_pi0d_diff} demonstrates huge spurious contribution of
the coherent process in IA which is eliminated by applying the
modified IA. This result could be expected because the effect 
of orthogonality should be especially
visible in the region of small pion angles, where the momentum transferred to
the nucleons is minimal and the overlap between initial and final wave
functions is most important.
The additional FSI effects are quite small. Only at
350~MeV one notes a slight increase of the cross section near the maximum
around 60$^\circ$ and at 780~MeV a more sizeable increase.

\subsection{Beam asymmetry for linearly polarized photons}

\begin{figure}[htb]
\includegraphics[scale=.6]{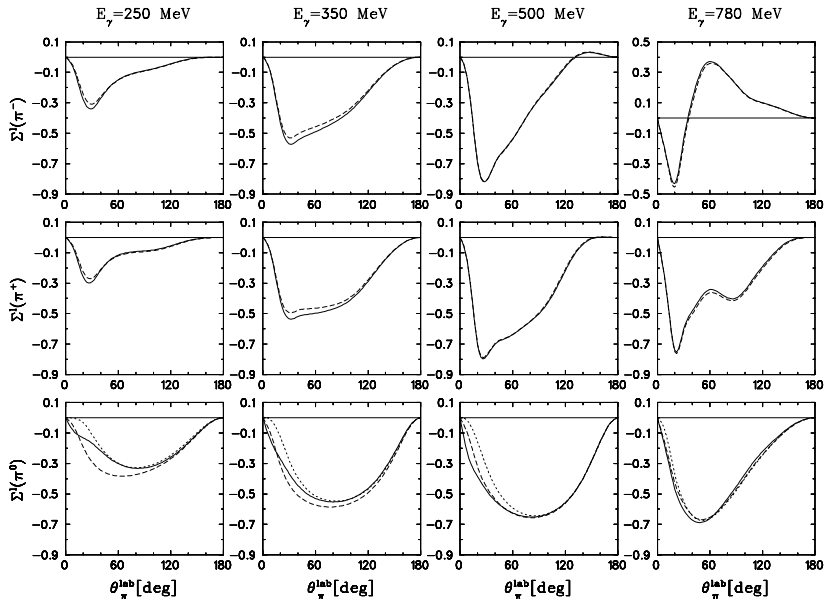}
\caption{Photon asymmetry $\Sigma^l$ for the three charge states of
semi-exclusive single pion photoproduction on the deuteron. Notation
of curves: dashed: IA; dotted: modified IA for
\protect$\pi^0$-production; solid: IA + \protect$NN$-rescattering.} 
\label{fig_sigma}
\end{figure}

As next we turn to the beam asymmetry $\Sigma^l$ displayed in
Fig.~\ref{fig_sigma} for the three charge states at various energies
in the $\Delta$-resonance region and above. In all of these
asymmetries one notes a relatively small influence from FSI. We will
consider first the results for $\pi^-$-production in the top panels of
Fig.~\ref{fig_sigma}. For the lowest three energies the photon
asymmetry $\Sigma^l$ is negative. It is quite small below the
$\Delta$-region and increases considerably in absolute size with
increasing energy, becoming quite broad around 350~Mev but more
forward peaked at 500~Mev. At the highest energy of 780~MeV one notes
a different behavior. The deep minimum in forward direction turns into
a broad positive distribution above 40$^\circ$. The three lowest
energies are also representative for $\pi^+$-production (middle panels
of Fig.~\ref{fig_sigma}). However at the highest energy the asymmetry
remains negative, but exhibits a secondary minimum around 90$^\circ$.

The $\Sigma^l$-asymmetry for $\pi^0$-production, shown in the lowest
panels of Fig.~\ref{fig_sigma}, exhibits quite a different behavior
compared to charged pion production. One finds a broad, structureless
and sinus-shaped negative distribution with a maximum around 80$^\circ$ to
90$^\circ$ with increasing amplitude by about a factor of two going
from 250 to 500~MeV. At 780~MeV the width of the 
distribution becomes smaller and
the minimum moves towards forward angles. In general also here the
influence of rescattering is small. It is mainly due to the removal of the
spurious coherent contribution. The relatively largest influence
appears at the lowest energy of $E_\gamma=250$~MeV.  

Compared to the results in~\cite{Dar04a,Dar05a,Dar05c} one
readily notes quite substantial differences for both, charged and
neutral pion production, both in angular behavior and also in absolute
size, in particular at higher energies. For $\pi^-$-production a
larger influence of FSI is found in~\cite{DaS05} which we cannot
confirm. 
Furthermore, $\Sigma^l$ does not vanish in~\cite{DaS05} at $\theta_q=0$
and $\pi$ as it should, although it is small. Strangely enough, the
authors mention this feature as a notable effect. The origin of these
differences is not clear. In any case, the comparison suffers from the
questionable formal expressions used for the calculation of $\Sigma^l$ 
which, moreover, differ in the various 
publications~\cite{Dar04a,Dar05c,DaS05}. 

\subsection{Target asymmetries for oriented deuteron}

\begin{figure}[htb]
\includegraphics[scale=.7]{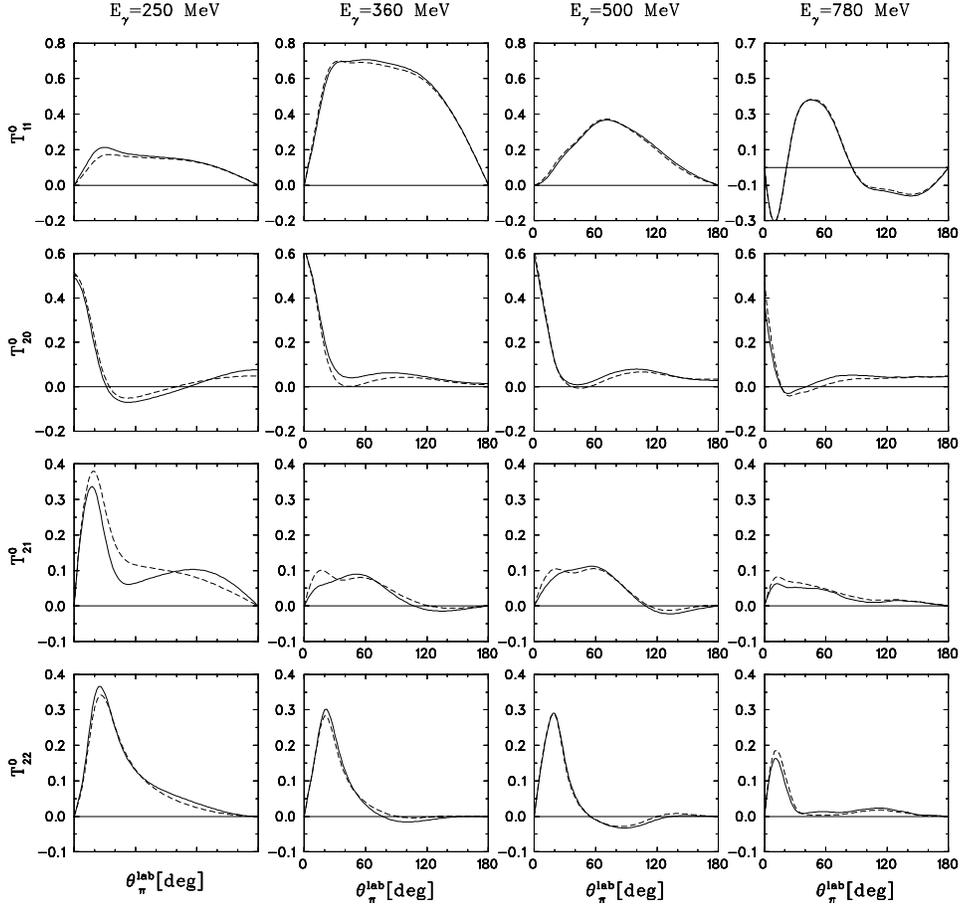}
\caption{Target asymmetries \protect$T^0_{IM}$ for semi-exclusive
\protect$\pi^-$-photoproduction on the deuteron. Notation of curves as
in Fig.~\protect\ref{fig_sigma}.} 
\label{fig_t_IM_mi}
\end{figure}
\begin{figure}[htb]
\includegraphics[scale=.8]{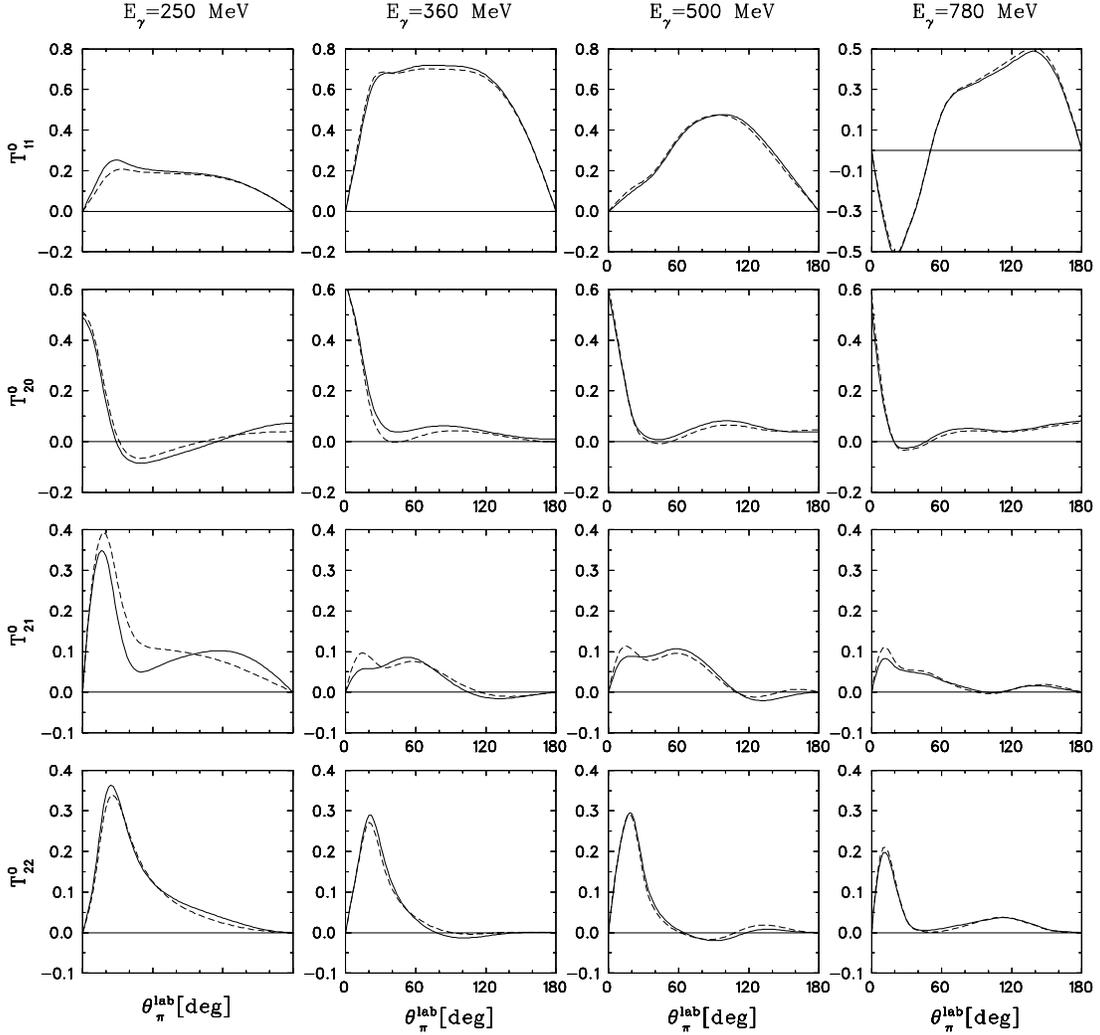}
\caption{Target asymmetries \protect$T^0_{IM}$ for semi-exclusive
\protect$\pi^+$-photoproduction on the deuteron. Notation of curves as
in Fig.~\protect\ref{fig_sigma}.} 
\label{fig_t_IM_pl}
\end{figure}
\begin{figure}[htb]
\includegraphics[scale=.8]{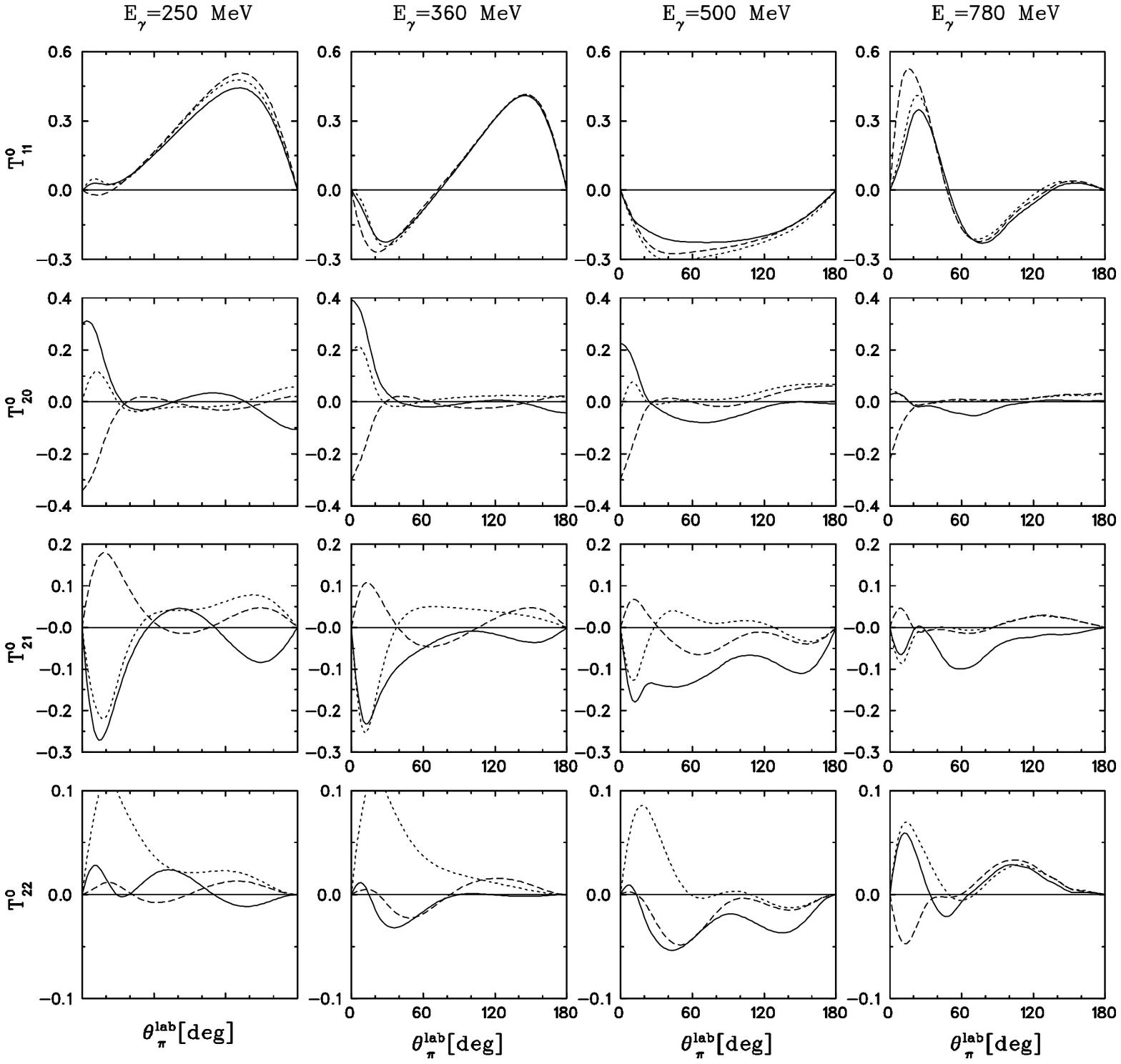}
\caption{Target asymmetries \protect$T^0_{IM}$ for semi-exclusive
\protect$\pi^0$-photoproduction on the deuteron. Notation of curves as
in Fig.~\protect\ref{fig_sigma}.}
\label{fig_t_IM_0}
\end{figure}

The target asymmetries $T^0_{IM}$ are shown in Figs.~\ref{fig_t_IM_mi}
through \ref{fig_t_IM_0} for the three charge states, respectively. 
In view of the similarity of the asymmetries for the two charge
states except for $T^0_{11}$ at 780~MeV, we will restrict the
discussion to $\pi^-$- and $\pi^0$-production. In general FSI effects 
are again small, the largest appear in $T^0_{21}$. The vector asymmetry 
$T^0_{11}$ is positive up to 500~MeV and shows a broad distribution
over the whole angular range. But at 780~MeV an oscillatory behavior
develops with sizeable amplitude. Only at the lowest energy of
$E_\gamma=250$~MeV one notes some notable influence at forward angles
from FSI. With respect to~\cite{Dar04a,Dar05a,DaS05} we find an
opposite sign, a smaller amplitude and also a somewhat different
angular behavior. 

The tensor target asymmetry $T^0_{20}$ exhibits a pronounced sharp
peak at 0$^\circ$ and a rapid fall-off with increasing angles,
remaining quite small above 30$^\circ$. The results are similar
to~\cite{Dar04a,Dar05a,DaS05} except that in contrast to the small
negative values of~\cite{Dar04a,Dar05a} at backward angles, we find
small positive values. Moreover, we find smaller FSI influences at
backward angles than in~\cite{DaS05}. $T^0_{21}$ peaks at small angles
around 20$^\circ$ for $E_\gamma=250$~MeV which disappears at the three
higher energies becoming a broader distribution with a considerably
smaller size. FSI shows some notable influence. Again we find
significant differences to~\cite{Dar04a,Dar05a} with respect to shape,
size, and FSI effects, in particular at the lowest energy. Finally,
$T^0_{22}$ exhibits a prominent peak in forward direction which
becomes sharper and moves towards smaller angles with increasing photon
energy. This is in qualitative agreement with~\cite{Dar04a,Dar05a}
although the size is different and our FSI effects are much smaller. 

The target asymmetries for $\pi^0$-production in Fig.~\ref{fig_t_IM_0}
show quite a different behavior compared to charged pion
production. The structure of $T^0_{11}$ changes significantly
with energy. While at $E_\gamma=250$~MeV one finds a maximum
around 130$^\circ$, one notes a forward negative minimum and a
backward positive maximum at $E_\gamma=350$~MeV, at 
$E_\gamma=500$~MeV a broad and quite flat negative distribution, and
finally at 780~MeV a forward maximum and a negative minimum around
80$^\circ$. The tensor asymmetries are much more sensitive to
FSI. This is particularly apparent in $T^0_{20}$ exhibiting a forward
negative minimum in IA which turns into a positive forward peak when
FSI is switched on. Also $T^0_{21}$ shows such a drastic influence
from FSI. $T^0_{22}$ is much smaller than for charged pion production
and shows an oscillatory behavior. FSI effects are noticeable
again. For these asymmetries the differences to the results
of~\cite{Dar04a,Dar05a} are again quite significant. 

\subsection{Beam-target asymmetries for circularly polarized photons}

As next we will discuss the double polarization asymmetries $T_{IM}^c$
for circularly polarized photons and oriented deuterons. They are 
in a certain sense complementary to the target asymmetries $T^0_{IM}$
because, while one is the real part, the other is the imaginary part of
the basic quantities $V_{IM}^1$ as defined in~\cite{ArF05}. Since the
results for the two charged pions are again in general quite similar,
we display the results only for $\pi^-$ in
Fig.~\ref{fig_double_c_pimin} and for $\pi^0$ 
in Fig.~\ref{fig_double_c_pi0}. In contrast to what has been claimed 
in~\cite{Dar05a}, all of them are nonvanishing. As a sideremark,
although $T_{10}^c$ should vanish according to~\cite{Dar05a}, a
non-vanishing spin asymmetry is reported in~\cite{Dar05b}. For both,
$\pi^0$ and 
$\pi^-$, the vector asymmetry $T_{10}^c$ is quite sizeable in forward
and backward direction and also around 90$^\circ$ for $\pi^0$. It is
this asymmetry which determines the Gerasimov-Drell-Hearn sum
rule. The influence of FSI is quite marginal, in particular for
$\pi^-$. For both charge states the energy dependence is weak. The
other vector asymmetry $T_{11}^c$ shows a rather different
behavior. For $\pi^-$ it is predominantly negative with a slight
preference of the backward direction, especially at higher energies,
while for $\pi^0$ it has a positive maximum around 30$^\circ$ and a
negative minimum around 150$^\circ$, almost independent of the
energy. Considerably smaller are the tensor asymmetries $T_{21}^c$ and
$T_{22}^c$, but they are a little more sensitive to FSI. This is
particularly apparent for $T_{21}^c$ in $\pi^0$ production above the
$\Delta$-resonance region. 

\subsection{Beam-target asymmetries for linearly polarized photons}

The beam-target asymmetries $T_{IM}^l$ for linearly polarized photons
and polarized deuterons for $\pi^-$ and $\pi^0$
production are shown in Figs.~\ref{fig_double_l_pimin} through
\ref{fig_double_l_pi01}. Also these asymmetries are very similar for
the two charge states $\pi^-$ and $\pi^+$. The vector asymmetries
$T_{1M}^l$ do not vanish at all as claimed in~\cite{Dar04a}. They are
small for $\pi^-$ at the lowest energy of $E_\gamma=250$~MeV, but
become sizeable for the higher energies 
(see Fig.~\ref{fig_double_l_pimin}). They are considerably smaller for
$\pi^0$ production as shown in Fig.~\ref{fig_double_l_pi0} but more
sensitive to FSI. Of the corresponding tensor asymmetries in
Figs.~\ref{fig_double_l_pimin1} and \ref{fig_double_l_pi01},
$T_{22}^l$ is by far the largest for both charge states $\pi^-$ and
$\pi^0$, exhibiting a pronounced forward peak which becomes slightly
sharper with increasing energy. The size decreases significantly when
going from $T_{22}^l$ to $T_{2-2}^l$. By the way, they are certainly
not equal as claimed in~\cite{Dar04a}. While $T_{2-2}^l$ is restricted
to vanish at $\theta_q=0$ and $\pi$, this is not the case for
$T_{22}^l$. For $\pi^-$ production one notes 
some FSI effects in $T_{2\pm 1}^l$, whereas for $\pi^0$ quite drastic
influences from FSI can be seen, leading even to a sign change in some
of them. A comparison to the results in~\cite{Dar04a} makes no sense
because of the above mentioned wrong formal expressions in~\cite{Dar04a}.

\subsection{Comparison with experiment}

We now will turn to a comparison with experimental data, where
available. Fig.~\ref{fig_tot_exp} shows the total cross sections
for $\pi^-$- and $\pi^0$-production. The agreement with experimental data for
$\pi^-$-production is satisfactory although not perfect. The theory is
a little high in the maximum but low in the dip
region between the $\Delta$ and the second resonance region. For
$\pi^0$, one notes some slight overshooting in the maximum and a
sizeable overestimation above the $\Delta$ in the second resonance region. 
This discrepancy was already discussed in~\cite{Kru99}
where it was noted that the smearing and damping of the second
resonance peak can hardly be explained by the Fermi motion effect alone. 
According to our results, the broadening of the
resonance structure in pion production on the deuteron is quite
significant. This effect is readily seen in
Fig.~\ref{fig_pi0d_gN_enrg}, where we present differential cross
sections for $\pi^0$-production on the deuteron as functions of the
photon energy for fixed pion angle. The cross section refers to
the equivalent $\gamma N$ c.m.\ system, where the nucleon is at rest 
in the deuteron. The solid curves are our full
calculation multiplied by energy independent factors, whose values are
listed in the various panels. One readily sees that the Fermi motion and
to some extent FSI leads to a disappearing of the resonance structure. The
ratio of the theoretical cross section to the experimental result is about 
1.75, except for $\cos\theta_{\pi^0}^*=-0.9$ where it is about 3. However the
origin of this factor is unclear. 

One possible source could be 
the neglected interaction of nucleon resonances with the spectator 
nucleon which might result in a broadening of the resonances due to 
additional inelastic processes. In fact, within a simple model
calculation for the $\Delta$-resonance a significant lowering of the
total cross section from such an interaction was found in~\cite{ReA05}. 
A further source could be the neglected inelasticity of the final 
$NN$-interaction at higher energies. On the other hand the question 
arises, why the same lowering is not observed in the 
$\pi^-$-photoproduction where the agreement with the data is quite 
satisfactory.

The threshold region is shown in Fig.~\ref{fig_tot_threshold_exp} for
$\pi^+$-production. Good agreement with the data is achieved,
comparable in quality to a recent precision calculation in chiral
perturbation theory~\cite{LeB05}. The crucial role of $NN$-FSI at 
low energies, leading to a strong enhancement over the IA, is furthermore
demonstrated by the differential cross section with respect to
the relative energy of the two final neutrons in $\pi^+$-production 
in Fig.~\ref{fig_pipl_threshold} for two
different kinematical situations, pion emission in more forward (left
panel) and in more backward direction (right panel). The $^1S_0$-state
of $nn$-scattering near threshold, a manifestation of the so-called
antibound state as companion to the deuteron, is nicely resolved and
also quite well reproduced by the theory if $NN$-rescattering is
included. A similar result has been reported in~\cite{LeC04}. 

With respect to differential cross sections, a comparison for 
$\pi^-$-production is exhibited in Fig.~\ref{fig_piMi_diff_exp} 
and for $\pi^0$-production in Fig.~\ref{fig_pi0_diff_exp}. For 
$\pi^-$-production one notes a satisfactory agreement with 
experimental data from Benz et al.~\cite{Be+73}, whereas for 
$\pi^0$-production one finds for the three highest energies a 
slight overestimation of the theory in the maximum and at forward angles 
an underestimation. 

A comparison to recent data from the LEGS-collaboration
(LEGS-exp.L3b)~\cite{Sandorfi} on the unpolarized
semi-exclusive differential cross section and $\Sigma^l$ for 
$\pi^-$ production is shown in Fig.~\ref{fig_pimi_sig} for two
energies near the $\Delta$-region. One notes very little
influence from FSI and quite a good agreement of the theoretical
description with the data. Our results on $\Sigma^l$ are similar
to the IA calculation of Lee and Sato~\cite{Lee}, 
but at variance to~\cite{Dar05c} where also larger FSI
effect were quoted. But the latter results are questionable for
reasons mentioned above. Another comparison for $\Sigma^l$ with data from
the LEGS-collaboration~\cite{Sandorfi} 
are exhibited in Fig.~\ref{fig_pimi_sig_e} for
constant pion emission angle as function of the photon energy. In this
case the size is underestimated by the theory, although FSI shifts the
results for IA in the right direction but not enough. A much better
almost perfect agreement with respect to $\Sigma^l$ is
shown in Fig.~\ref{fig_pi0_sig} for $\pi^0$-production.

Finally, we show in Fig.~\ref{fig_piMi_GDH} for $\pi^-$-production
a comparison between theory and experiment for the
semi-exclusive differential spin asymmetry
$d^2(\sigma^P-\sigma^A)/d\Omega_q$ 
with respect to circularly polarized photons and the deuteron spin oriented
parallel (P) or antiparallel (A) to the photon spin. This spin asymmetry 
is related to the beam-target asymmetry $T^c_{10}$ according to
\beq
\frac{d^2(\sigma^P-\sigma^A)}{d\Omega_q}
=\sqrt{6}\,\frac{d^2\sigma_0}{d\Omega_q}\,T^c_{10}\,.
\eeq
Compared to the predictions in~\cite{Dar05b} we find the depth of the 
minimum at 0$^\circ$ almost independent of the energy and a small positive
asymmetry between 30$^\circ$ and 60$^\circ$. This spin asymmetry has been 
measured by the A2-collaboration~\cite{Pedroni}. However, the analysis of 
the data is not yet completed. Preliminary data were shown in~\cite{Dar05b}
without authorization and thus are not shown here. Compared to these,
the agreement is quite satisfactory, although in the 
angular region, where the data are available, the spin asymmetry is very
small, almost compatible with zero. Thus it would be very desirable to
have additional data at more forward angles, where the theoretical 
asymmetry exhibits a pronounced minimum. This concludes the discussion 
of results.

\section{Conclusion and outlook}\label{outlook}

In the present work we have exploited the role of polarization
observables in incoherent pion photoproduction on the deuteron with
particular emphasis on the influence of final state interaction in the
$NN$- and $\pi N$-subsystems of the final state. In the unpolarized total 
and semi-exclusive differential cross section $d^2\sigma/d\Omega_q$, where 
only the direction of the produced pion is measured, the influence of 
final state rescattering is quite small for charged pion production for 
photon energies up to 1~GeV. For $\pi^0$-production the influence is 
much larger. However, the dominant part of FSI-effect arises from the 
removal of a spurious coherent contribution in the impulse approximation 
when $NN$-rescattering is switched on. This is demonstrated by a modified 
IA, where the deuteron wave function component in the final $NN$-plane 
wave is projected out. The remaining FSI-effect is comparable to charged 
pion production. 

As polarization observables we have considered all beam, target and 
beam-target asymmetries of the semi-exclusive differential cross 
section. Many of them are quite sizeable, in particular the photon 
asymmetry $\Sigma^l$ and the various vector asymmetries. The tensor
asymmetries are in general considerably smaller. They are often
quite insensitive to final state rescattering. Only a few, $T_{21}$, and
$T_{21}^c$, show a larger influence in charged pion production. 

A very interesting and still open question concerns the disagreement between
theoretical and experimental results for $\gamma d\to\pi^0 np$ in the second
resonance region. Although our calculation explains the strong smearing of the
resonance structure, the data are overestimated by about a factor of
1.5 (see Figs.~\ref{fig_tot_exp} and \ref{fig_pi0d_gN_enrg}). 
Hopefully, new measurements of the ratio 
\begin{equation}
R=\frac{d\sigma(\gamma,\pi^0n)}{d\sigma(\gamma,\pi^0p)}
\end{equation}
for quasifree photoproduction on neutrons and protons can clarify the
situation. The old data from \cite{Bacci,Hemmi}, 
pointing to $R=1$ in the second resonance region, seem to be 
in disagreement with the results of \cite{Kru99}.
 
Future theoretical improvements should be devoted to the inclusion 
of two-body effects
in the photoproduction amplitude, e.g.\ the interaction between a resonance 
and the spectator nucleon, inclusion of inelasticities in the $NN$-interaction
and the role of relativistic effects at higher energies. The problem
of off-shell effects for the elementary amplitude is another unsolved task. 
Furthermore, there is urgent need for a unified description of single 
and double pion production. 

\acknowledgments                                                      
We would like to thank Michael Schwamb for interesting discussions and a
careful reading of the manuscript.
This work was supported by the Deutsche Forschungsgemeinschaft (SFB 443).

\appendix
\renewcommand{\theequation}{A\arabic{equation}}
\setcounter{equation}{0}
\section{Separation of polarization asymmetries}
\label{appa}
In this appendix we will discuss how the various polarization
asymmetries of the semi-incusive differential cross section can be
separated by a proper variation of the photon polarization parameters
($P^\gamma_l$ and $P^\gamma_c$), the deuteron polarization parameters
($P_1^d$ and $P_2^d$), the polarization angles ($\theta_d$ and
$\phi_d$), and the dependence on the pion azimuthal angle $\phi_q$,
similar to what has been described in~\cite{ArL05}. To this end we
write the differential cross section of (\ref{diffcrossc}) as follows 
\beq
S(P^\gamma_l,P^\gamma_c,P^d_1,P^d_2)=S_0
\Big[1+ P^d_1\,A_{d}^{V} + P^d_2\,A_{d}^{T} 
+P^\gamma_l\,(A^l_\gamma + P^d_1\,A_{\gamma d}^{lV}+ 
P^d_2\,A_{\gamma d}^{lT})+
P^\gamma_c\,(P^d_1\,A_{\gamma d}^{cV}+ P^d_2\,A_{\gamma d}^{cT})
\Big]\,,\label{diffcross_sep}
\eeq
with $S_0=S(0,0,0,0)$ as unpolarized differential cross
section. Furthermore, we have introduced as generalized
single polarization $A$-asymmetries  
\beqa
A_{d}^{V}(\theta_q,\phi_{qd},\theta_d)&=&
\widetilde T_{11}^0(\theta_q)\,\sin\phi_{qd}\,d^1_{10}(\theta_d)\,,
\label{adv}\\
A_{d}^{T}(\theta_q,\phi_q,\phi_{qd},\theta_d)&=&
\sum_{M= 0}^2\widetilde T_{2M}^0(\theta_q)\cos(M\phi_{qd})
\,d^2_{M0}(\theta_d)\,,\\
A^l_\gamma(\theta_q,\phi_{q})&=&\widetilde \Sigma^l(\theta_q)\,\cos
2\phi_q\,,
\eeqa
and double polarization $A$-asymmetries 
\beqa
A_{\gamma d}^{lV}(\theta_q,\phi_q,\phi_{qd},\theta_d)&=&
\sum_{M= -1}^1\widetilde T_{1M}^l(\theta_q)\,
\sin(M\phi_{qd}-2\,\phi_q) \,d^1_{M0}(\theta_d)\,\nonumber\\
&=&\sum_{M= -1}^1\widetilde T_{1M}^l(\theta_q)\,
[\sin(M\phi_{qd})\cos(2\,\phi_q)-\cos(M\phi_{qd})\,\sin(2\,\phi_q)]
\,d^1_{M0}(\theta_d)\,,\\
A_{\gamma d}^{lT}(\theta_q,\phi_q,\phi_{qd},\theta_d)&=&
\sum_{M= -2}^2\widetilde T_{2M}^l(\theta_q)\,
\cos(M\phi_{qd}-2\,\phi_q)\,d^2_{M0}(\theta_d)\,\nonumber\\
&=&\sum_{M= -2}^2\widetilde T_{2M}^l(\theta_q)\,
[\cos(M\phi_{qd})\,\cos(2\,\phi_q)+\sin(M\phi_{qd})\,\sin(2\,\phi_q)]
\,d^2_{M0}(\theta_d)\,,\\
A_{\gamma d}^{cV}(\theta_q,\phi_{qd},\theta_d)&=&
\sum_{M= 0}^1\widetilde T_{1M}^c(\theta_q)\,
\cos(M\phi_{qd})\,d^1_{M0}(\theta_d)\,,\\
A_{\gamma d}^{cT}(\theta_q,\phi_{qd},\theta_d)&=&
\sum_{M= 1}^2\widetilde T_{2M}^c(\theta_q)\,
\sin(M\phi_{qd})\,d^2_{M0}(\theta_d)\,.\label{agdct}
\eeqa
Each of these generalized $A$-asymmetries are functions of $\theta_q$ and
of some of the variables $\phi_q$, $\phi_{qd}$, and
$\theta_d$. The first step is to isolate them
by evaluating linear combinations of cross sections for different
values of the appropriate polarization parameters 
($P^\gamma_{l,c}$, $P^d_{1,2}$). For example,
$A^l_\gamma$ is obtained from the cross section difference for
unpolarized deuterons and linearly polarized photons ($P^\gamma_l>0$)
minus the unpolarized one ($P^\gamma_l=0$), i.e.\ 
\beq
A^l_\gamma=\frac{1}{P^\gamma_l\,S_0}\,
\Big[S(P^\gamma_l,0,0,0)-S_0\Big]\,.
\eeq
The others can be obtained by the following linear combinations of
cross sections
\beqa
A^V_d&=&\frac{1}{2\,P^d_1\,S_0}\,
\Big[S(0,0,P^d_1,P^d_2)-S(0,0,-P^d_1,P^d_2)\Big]\,,\\
A^T_d&=&\frac{1}{2\,P^d_2\,S_0}\,
\Big[S(0,0,P^d_1,P^d_2)+S(0,0,-P^d_1,P^d_2)-2\,S_0\Big]\,,\\
A^{cV}_{\gamma d}&=&\frac{1}{4\,P^\gamma_c\,P^d_1\,S_0}\,
\Big[S(0,P^\gamma_c,P^d_1,P^d_2)-S(0,-P^\gamma_c,P^d_1,P^d_2)
-S(0,P^\gamma_c,-P^d_1,P^d_2)+S(0,-P^\gamma_c,-P^d_1,P^d_2)\Big]\,,\\
A^{cT}_{\gamma d}&=&\frac{1}{4\,P^\gamma_c\,P^d_2\,S_0}\,
\Big[S(0,P^\gamma_c,P^d_1,P^d_2)-S(0,-P^\gamma_c,P^d_1,P^d_2)
+S(0,P^\gamma_c,-P^d_1,P^d_2)-S(0,-P^\gamma_c,-P^d_1,P^d_2)\Big]\,,\\
A^{lV}_{\gamma d}&=&\frac{1}{2\,P^\gamma_l\,P^d_1\,S_0}\,
\Big[S(P^\gamma_l,0,P^d_1,P^d_2)-S(0,0,P^d_1,P^d_2)
-S(P^\gamma_l,0,-P^d_1,P^d_2)+S(0,0,-P^d_1,P^d_2)\Big]\,,\\
A^{lT}_{\gamma d}&=&\frac{1}{2\,P^\gamma_l\,P^d_2\,S_0}\,
\Big[S(P^\gamma_l,0,P^d_1,P^d_2)-S(0,0,P^d_1,P^d_2)
+S(P^\gamma_l,0,-P^d_1,P^d_2)-S(0,0,-P^d_1,P^d_2)\nonumber\\&&
-2\,P^\gamma_l\,A^l_\gamma\,S_0\Big]\,.
\eeqa
Two of the generalized $A$-asymmetries in (\ref{adv}) through (\ref{agdct}) 
yield directly one asymmetry each,
namely $\widetilde \Sigma^l$ from $A^l_\gamma$ and $\widetilde
T_{11}^0$ from $A_{d}^{V}$, i.e.\ 
\beqa
\widetilde \Sigma^l&=&A^l_\gamma({\phi_q=0})\,,\\
\widetilde T_{11}^0&=&
\sqrt{2}\,A_{d}^{V}({\phi_{qd}=\pi/2,\,\theta_d=\pi/2})\,.
\eeqa
One could also obtain $\widetilde \Sigma^l$ from taking for linearly
polarized photons the cross section difference of pions in the photon
plane to pions perpendicular to this plane, i.e.\
\beq
\widetilde \Sigma^l=\frac{1}{P^\gamma_l}\,
\frac{S(P^\gamma_l>0,0,0,0)|_{\phi_q=0}
-S(P^\gamma_l>0,0,0,0)|_{\phi_q=\pi/2}}
{S(P^\gamma_l>0,0,0,0)|_{\phi_q=0}
+S(P^\gamma_l>0,0,0,0)|_{\phi_q=\pi/2}}\,.
\eeq
The remaining general asymmetries $A_d^T$ and $A_{\gamma
d}^{(l/c)(V/T)}$ contain linear combinations of the asymmetries
$\widetilde T_{2M}^0$ and $\widetilde T_{IM}^{l/c}$. In order to
separate the latter, one can exploit the dependence of the
$A$-asymmetries on the angular variables $\phi_q$, $\phi_{qd}$, and
$\theta_d$. This is achieved, following the analogous problem in deuteron
electrodisintegration~\cite{ArL05}, by observing that the general
functional form of an $A$-asymmetry is
\begin{equation}
A^I(\phi_q,\phi_{qd},\theta_d)=\sum_{M=-I}^I \alpha_{IM}(\phi_q, 
\phi_{qd})d_{M0}^I(\theta_d)\,,\quad (I=1,2)\,,\label{AI}
\end{equation}
where
\begin{equation}\label{alphaIM}
\alpha_{IM}(\phi_q, \phi_{qd})=c_{IM}(\phi_q) \cos M\phi_{qd} +
s_{IM}(\phi_q) \sin M\phi_{qd}\;,
\end{equation}
and the $\phi_q$-dependent functions $c_{IM}(\phi_q)$ and $s_{IM}(\phi_q)$ 
have either the form
\begin{subequations}\label{csIM}
\begin{equation}
a_0 + a_1 \cos \phi_q + a_2 \cos 2\phi_q \label{(34)}
\end{equation}
or
\begin{equation}
b_1 \sin \phi_q + b_2 \sin 2\phi_q\;.
\end{equation}
\end{subequations}
One should note, that for $A_{\gamma d}^{c(V/T)}$ the
$\phi_q$-dependence is absent, i.e.\ $c_{IM}\equiv 0$ or $s_{IM}\equiv
0$, and the sum over $M$ in (\ref{AI}) runs from 0 through $I$. 
For a given $I$ the $M$-components $\alpha_{IM}(\phi_q, \phi_{qd})$ 
of the asymmetry 
$A^I(\phi_q,\phi_{qd},\theta_d)$ can be separated 
by a proper choice of $\theta_d$ exploiting the properties of the small
$d^I_{M0}$-functions. For $I=1$ (vector asymmetries), taking 
$\theta_d=0$ or $\pi /2$, i.e.\ $d^1_{M0}(0)=\delta_{M0}$ or 
$d^1_{M0}(\pi /2)=M/\sqrt{2}$, yields  $\alpha_{10}$ or $\alpha_{11}$, 
respectively, and for the tensor asymmetries ($I=2$) one may first choose 
$\theta_d=0$ yielding with $d^2_{M0}(0)=\delta_{M0}$ directly 
$\alpha_{20}$. The latter being determined, then setting $\theta_d={\pi /4}$ 
and $\pi /2$, one can obtain the remaining two terms $\alpha_{21}$ and 
$\alpha_{22}$. For the separation of $\alpha_{21}$ and $\alpha_{22}$ one 
can also choose $\theta_d=\theta_d^0=\mbox{arcos}\,(1/\sqrt{3})$ together with 
${\phi}_{qd}$ and ${\phi}_{qd}+\pi$. Then the sum and difference of the 
corresponding asymmetries result in $\alpha_{21}$ and $\alpha_{22}$, 
respectively.

In the next step, in order to separate the two contributions $c_{IM}$ 
and $s_{IM}$ in (\ref{alphaIM}), one can take first $\phi_{qd}=0$ 
giving $c_{IM}$ 
and then $\phi_{qd}={\pi / (2M)}$  for $M\neq 0$ which yields 
directly $s_{IM}$. 
The remaining separation of the coefficients $a_n$ or $b_n$ in 
(\ref{csIM}) is then achieved by appropriate 
choices of $\phi_q$. This completes the separation.

\renewcommand{\theequation}{B\arabic{equation}}
\setcounter{equation}{0}
\section{A modified impulse approximation}\label{appc}

For incoherent $\pi^0$-photoproduction in IA, the $NN$-final state 
is described by a plane wave $|\vec{p}, sm_s\rangle$ 
which is not orthogonal to the deuteron wave function. 
Thus the IA matrix element contains a spurious contribution from
coherent $\pi^0$-photoproduction, the size of which 
is governed by the overlap between the plane wave and the deuteron wave 
function which is just the deuteron wave function in momentum space
(see Eq.~(\ref{dwave})) 
\beq
\phi_{m_sm_d}(\vec{p}\,)=\langle \vec{p}, 1m_s|1m_d\rangle^{(d)}\ne 0\,,
\eeq
whereas, if the interaction between the nucleons is properly taken into 
account, the overlap with the final $NN$-scattering wave
vanishes. Therefore, one can expect that a large fraction of
$NN$-rescattering effects in incoherent $\pi^0$-photoproduction arises
from the elimination of the spurious coherent contribution. 

This spurious contribution can be avoided by applying a modified IA,
in which one uses a modified $NN$-final state wave function, where the
deuteron wave function component is projected out by the replacement 
\beq
|\vec{p},1m_s\rangle \rightarrow |\vec{p},1m_s\rangle 
- \sum_{m_d}|1m_d\rangle^{(d)}\langle 1m_d|\vec{p},1m_s\rangle=
|\vec{p},1m_s\rangle -
\sum_{m_d}|1m_d\rangle^{(d)}\,\phi_{m_sm_d}^*(\vec{p}\,)\,. 
\eeq
This means the following replacement for the IA matrix element
\beq
\langle \vec{p},1m_s|T|1m_d\rangle^{(d)} \rightarrow 
\langle \vec{p},1m_s|T|1m_d\rangle^{(d)} -
\sum_{m_d'}\phi_{m_sm_d'}(\vec{p}\,)\,^{(d)}\langle 
1m_d'|T|1m_d\rangle^{(d)}\,.
\eeq
The matrix element 
$^{(d)}\langle 1m_d'|T|1m_d\rangle^{(d)}$ corresponds to the one 
for coherent pion photoproduction $\gamma d\to\pi^0 d$ in the off-shell
region.
The comparison of the original IA with the modified IA reveals 
then what fraction of the whole FSI-effect arises from the non-orthogonality.

\begin{figure}[htb]
\includegraphics[scale=.8]{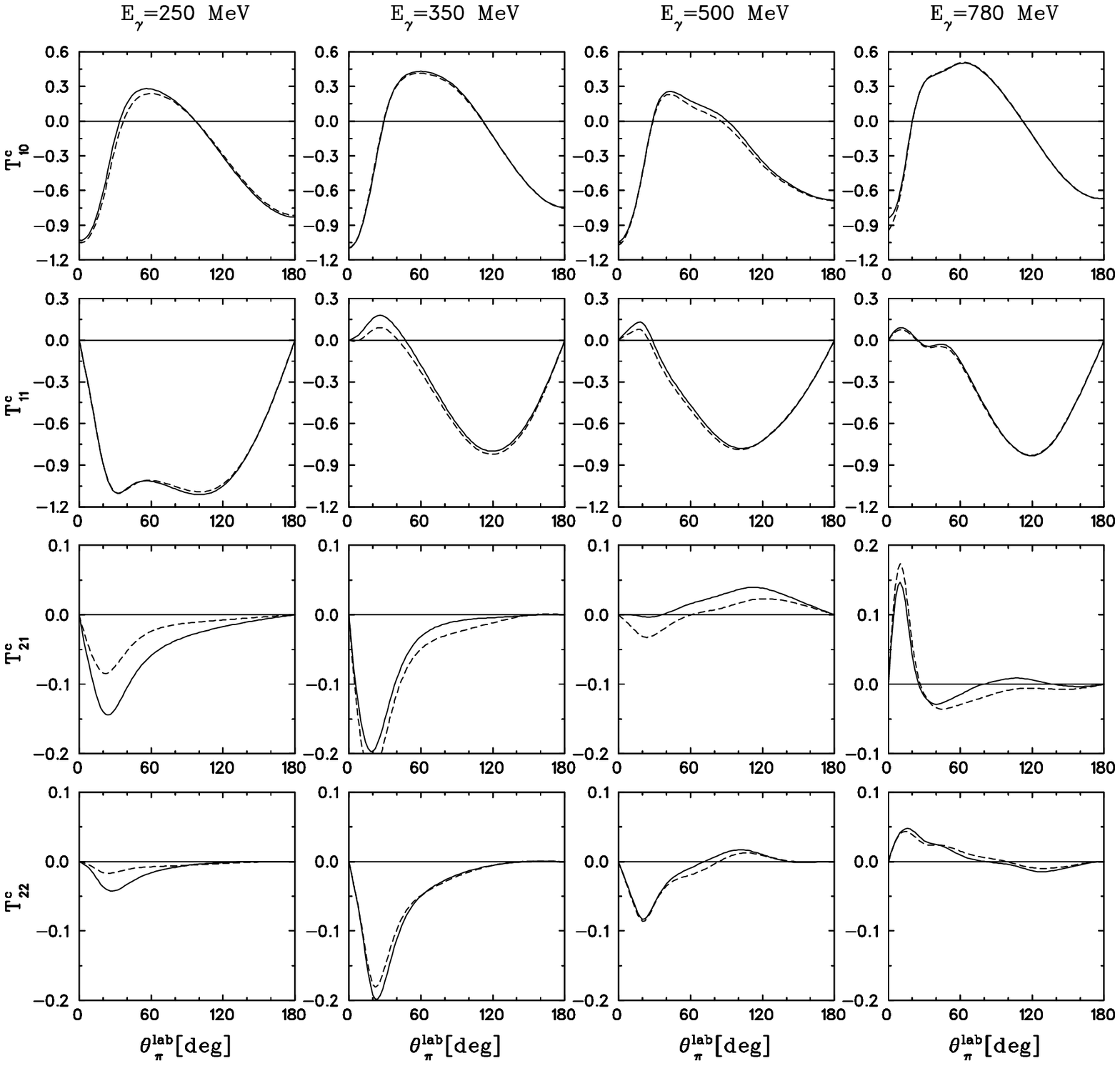}
\caption{Beam-target asymmetries \protect$T_{IM}^c$ for circularly
polarized photons and polarized deuterons for semi-exclusive
\protect$\pi^-$-photoproduction on the deuteron. Notation of curves as
in Fig.~\protect\ref{fig_sigma}.}
\label{fig_double_c_pimin}
\end{figure}

\begin{figure}[htb]
\includegraphics[scale=.8]{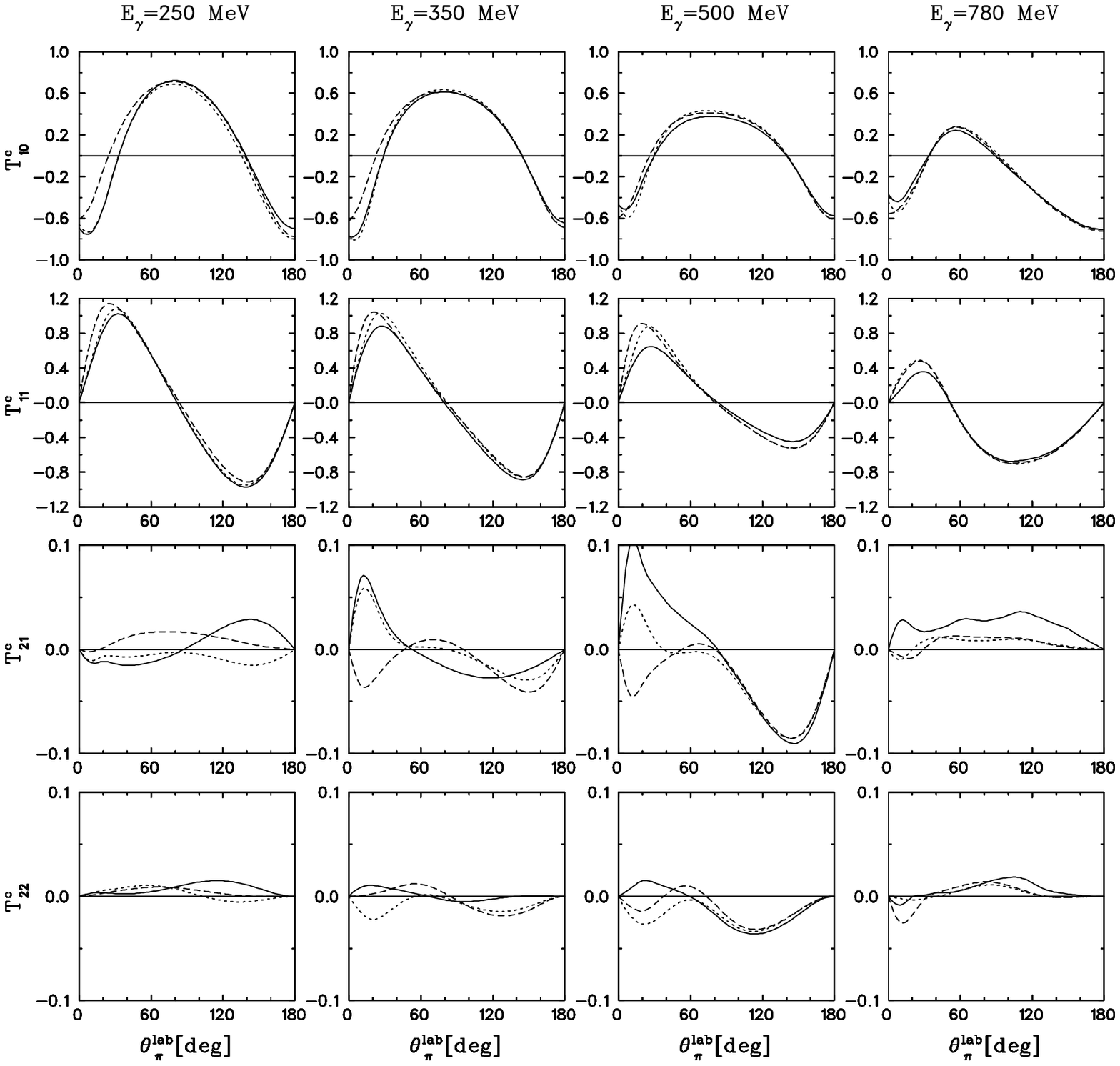}
\caption{Beam-target asymmetries \protect$T_{IM}^c$ for circularly
polarized photons and polarized deuterons for semi-exclusive
\protect$\pi^0$-photoproduction on the deuteron. Notation of curves as
in Fig.~\protect\ref{fig_sigma}.}
\label{fig_double_c_pi0}
\end{figure}

\begin{figure}[htb]
\includegraphics[scale=.7]{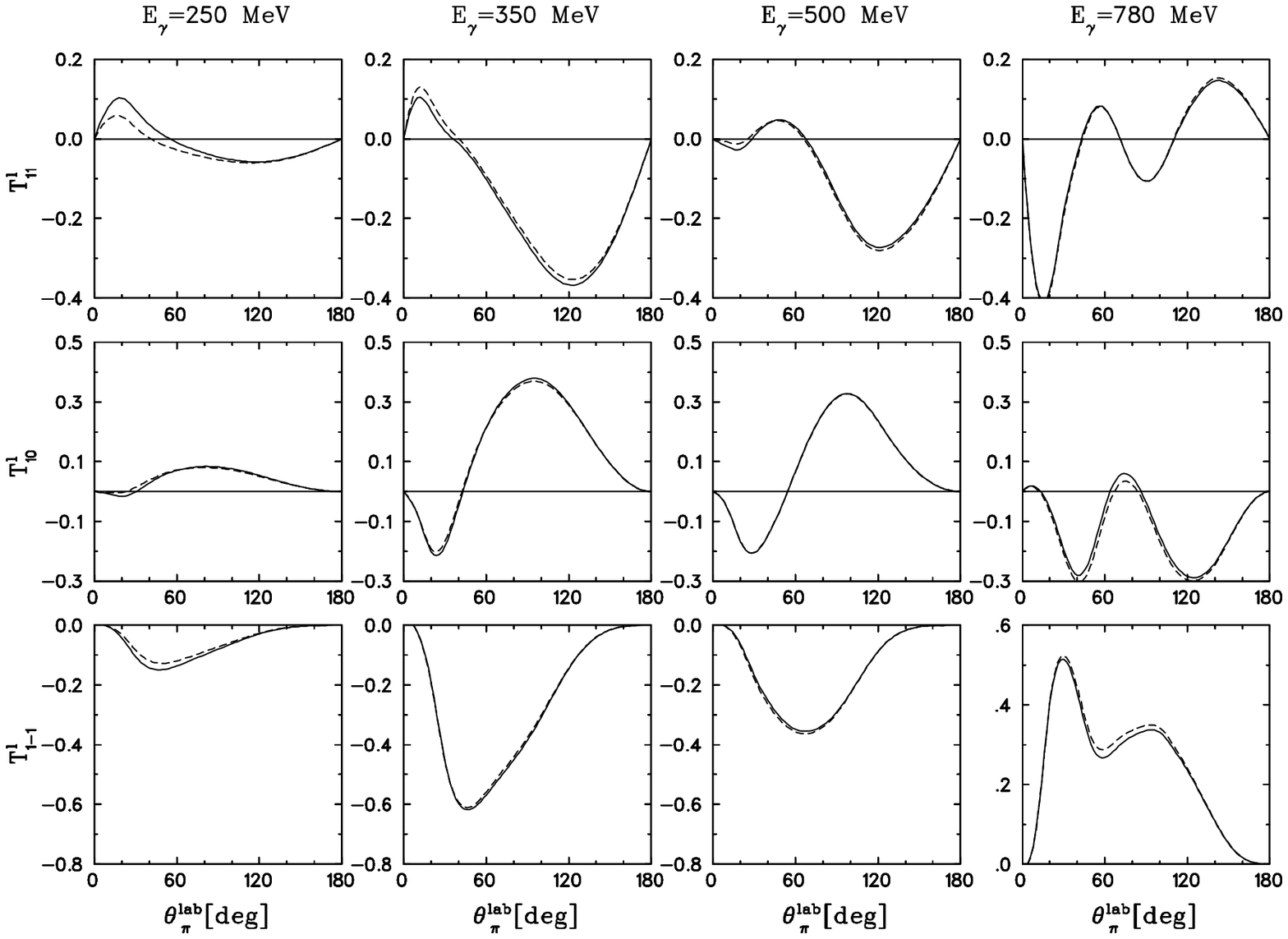}
\caption{Beam-target asymmetries \protect$T_{1M}^l$ for linearly
polarized photons and vector polarized deuterons for semi-exclusive
\protect$\pi^-$-photoproduction on the deuteron. Notation of curves as
in Fig.~\protect\ref{fig_sigma}.}
\label{fig_double_l_pimin}
\end{figure}

\begin{figure}[htb]
\includegraphics[scale=.7]{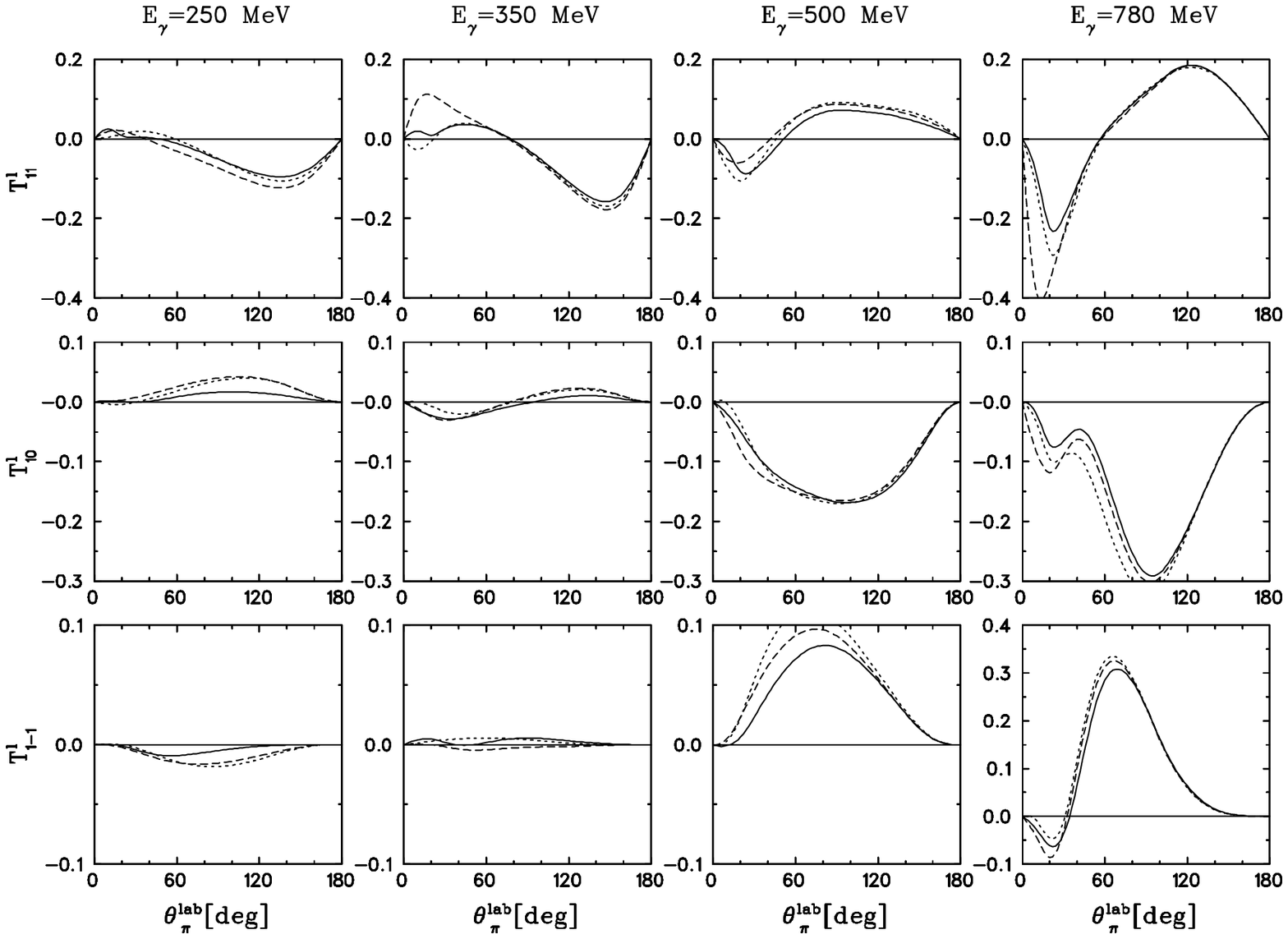}
\caption{Beam-target asymmetries \protect$T_{1M}^l$ for linearly
polarized photons and vector polarized deuterons for semi-exclusive
\protect$\pi^0$-photoproduction on the deuteron. Notation of curves as
in Fig.~\protect\ref{fig_sigma}.}
\label{fig_double_l_pi0}
\end{figure}

\begin{figure}[htb]
\includegraphics[scale=.8]{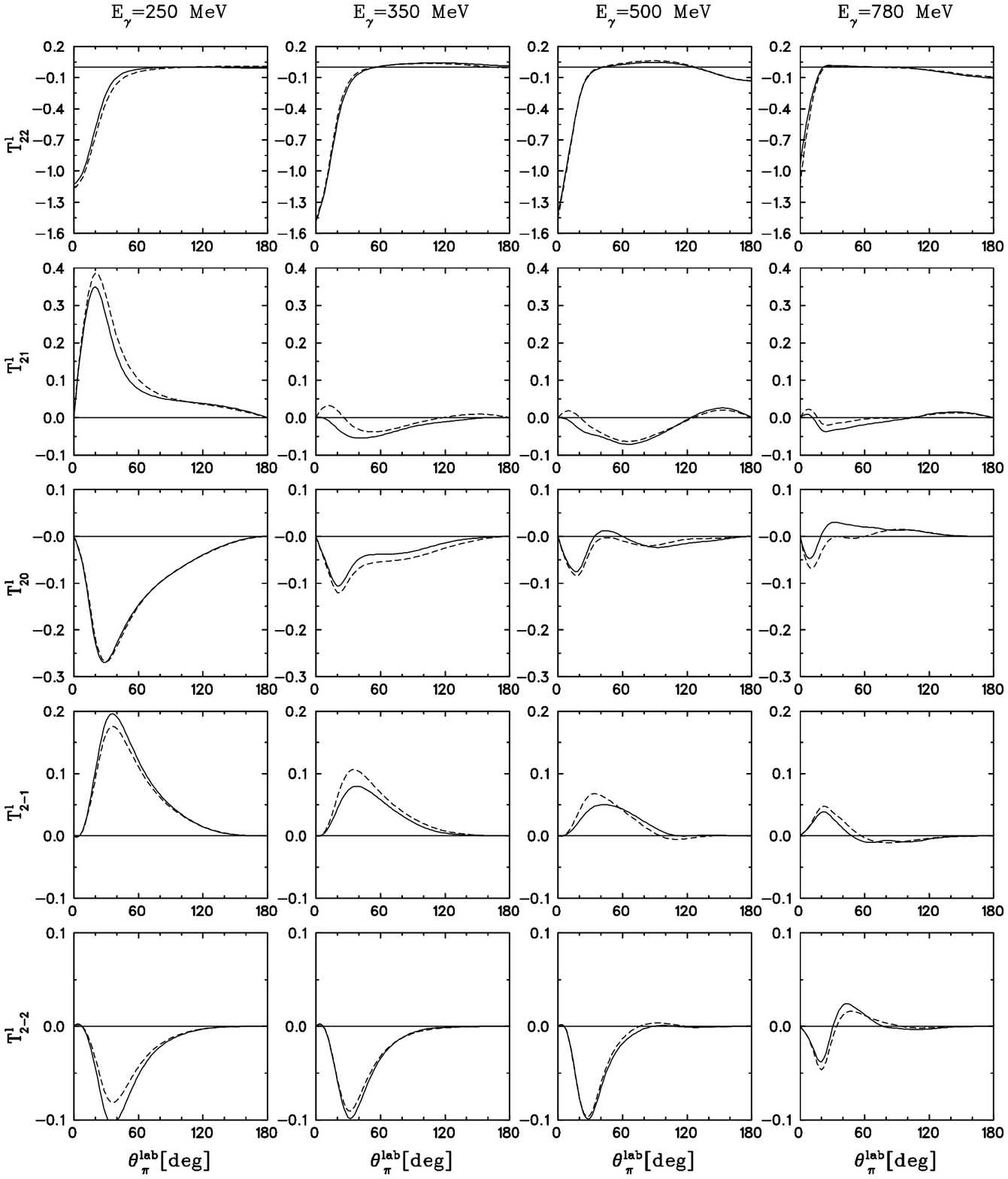}
\caption{Beam-target asymmetries \protect$T_{2M}^l$ for linearly
polarized photons and tensor polarized deuterons for semi-exclusive
\protect$\pi^-$-photoproduction on the deuteron. Notation of curves as
in Fig.~\protect\ref{fig_sigma}.}
\label{fig_double_l_pimin1}
\end{figure}

\begin{figure}[htb]
\includegraphics[scale=.8]{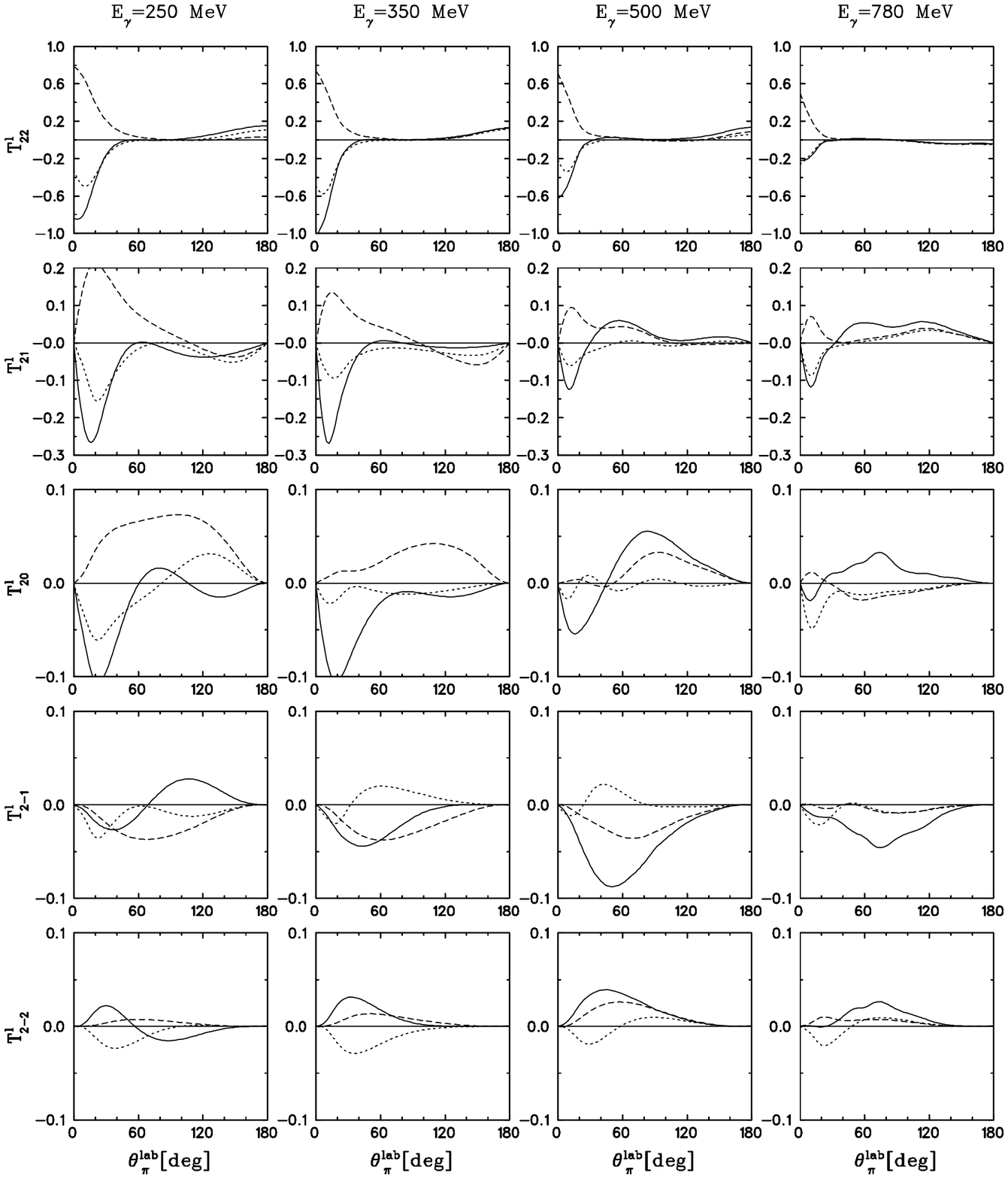}
\caption{Beam-target asymmetries \protect$T_{2M}^l$ for linearly
polarized photons and tensor polarized deuterons for semi-exclusive
\protect$\pi^0$-photoproduction on the deuteron. Notation of curves as
in Fig.~\protect\ref{fig_sigma}.}
\label{fig_double_l_pi01}
\end{figure}

\begin{figure}[htb]
\includegraphics[scale=.8]{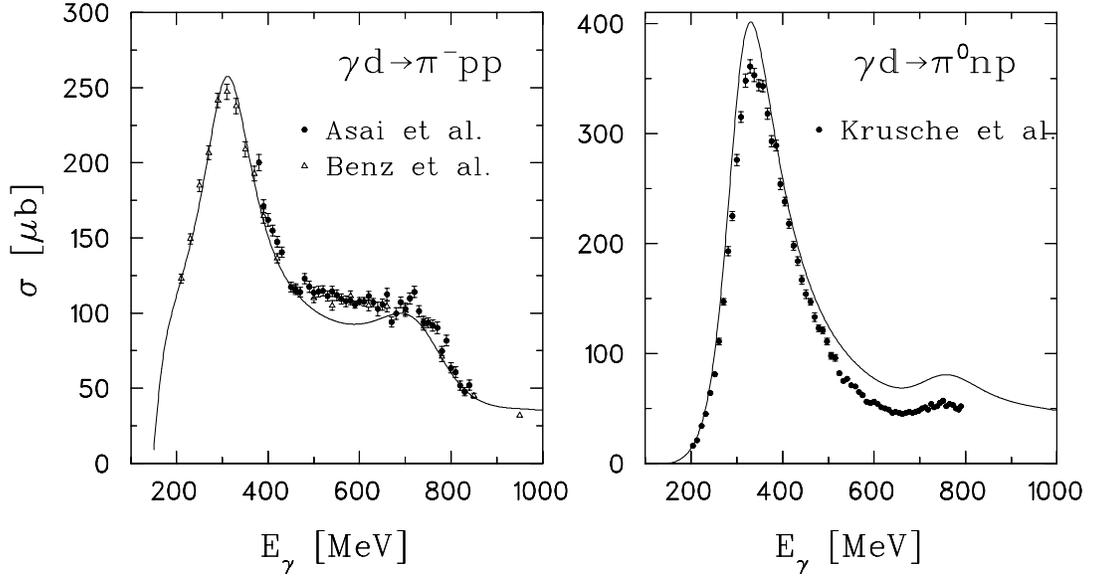}
\caption{Total cross section for \protect$\pi^-$ (left panel) and
\protect$\pi^0$-photoproduction (right panel) on the deuteron. Solid
curves: IA + \protect$NN$- and \protect$\pi N$-rescattering. 
Experimental data from Benz et
al.~\protect\cite{Be+73} and from Asai et al.~\protect\cite{As+90} for
$\pi^-$- and from Krusche et al.~\protect\cite{Kru99} for
$\pi^0$-production.} 
\label{fig_tot_exp}
\end{figure}

\begin{figure}[htb]
\includegraphics[scale=.7]{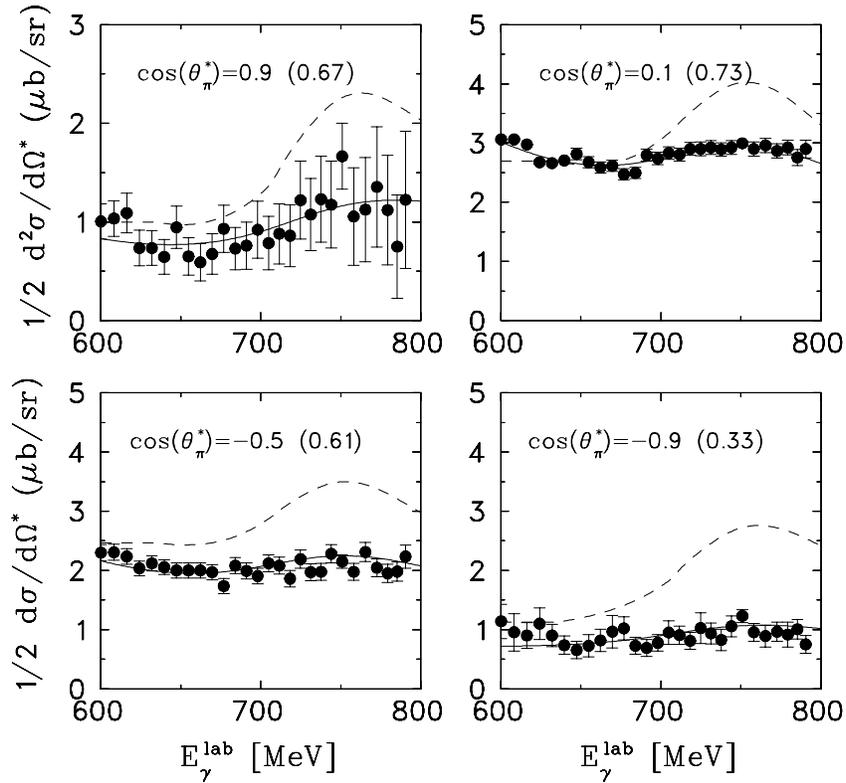}
\caption{Differential cross section for $\pi^0$-production on the
deuteron as function of the lab photon energy for fixed pion angles in
the equivalent $\gamma N$ c.m.\ system. The cross section is plotted
per nucleon. Notation of curves: solid: full calculation
multiplied by the factors in parentheses; dashed: MAID prediction for
the sum $1/2(d\sigma(\pi^0p)+d\sigma(\pi^0n))$.}
\label{fig_pi0d_gN_enrg}
\end{figure}

\begin{figure}[htb]
\includegraphics[scale=.9]{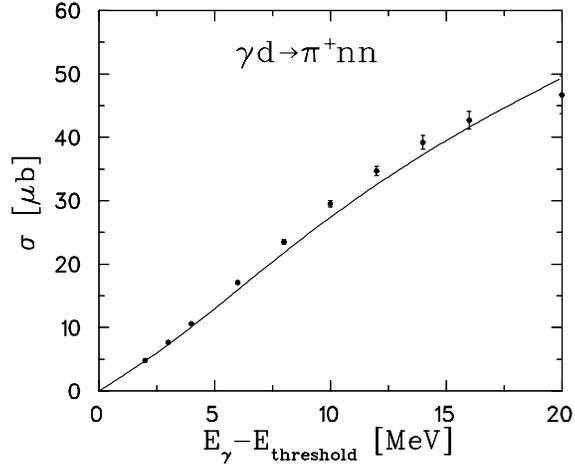}
\caption{Total cross section near threshold for
\protect$\pi^+$-photoproduction on the deuteron. Solid curve: IA +
\protect$NN$- and \protect$\pi N$-rescattering. Experimental data from
Booth et al.~\protect\cite{BoC79}.} 
\label{fig_tot_threshold_exp}
\end{figure}

\begin{figure}[htb]
\includegraphics[scale=.8]{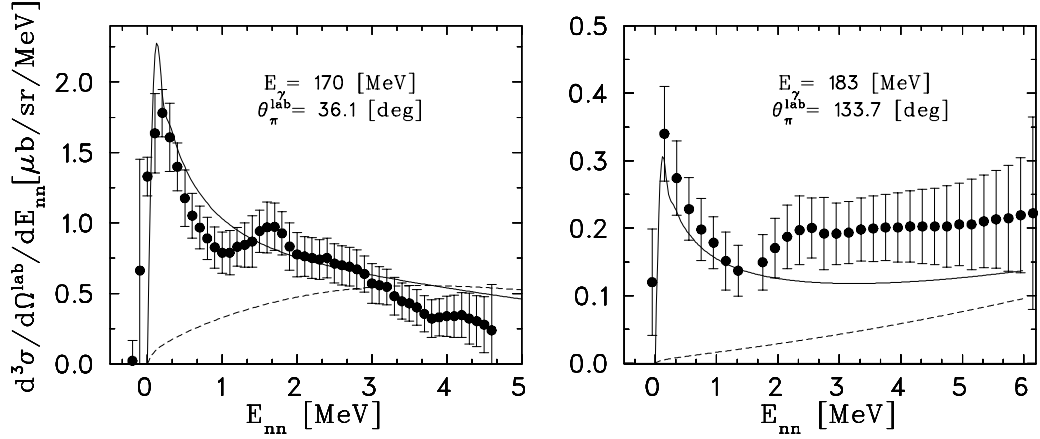}
\caption{Semi-exclusive differential cross section 
\protect$d^3\sigma/d\Omega_\pi\,dE_{nn}$ for $\pi^+$-photoproduction
on the deuteron at fixed pion angle as function of the relative energy
\protect$E_{nn}$ of the final two neutrons. Solid curves: IA +
\protect$NN$-rescattering. Experimental data from K\"obschall et
al.~\protect\cite{KoA87}.} 
\label{fig_pipl_threshold}
\end{figure}

\begin{figure}[htb]
\includegraphics[scale=.7]{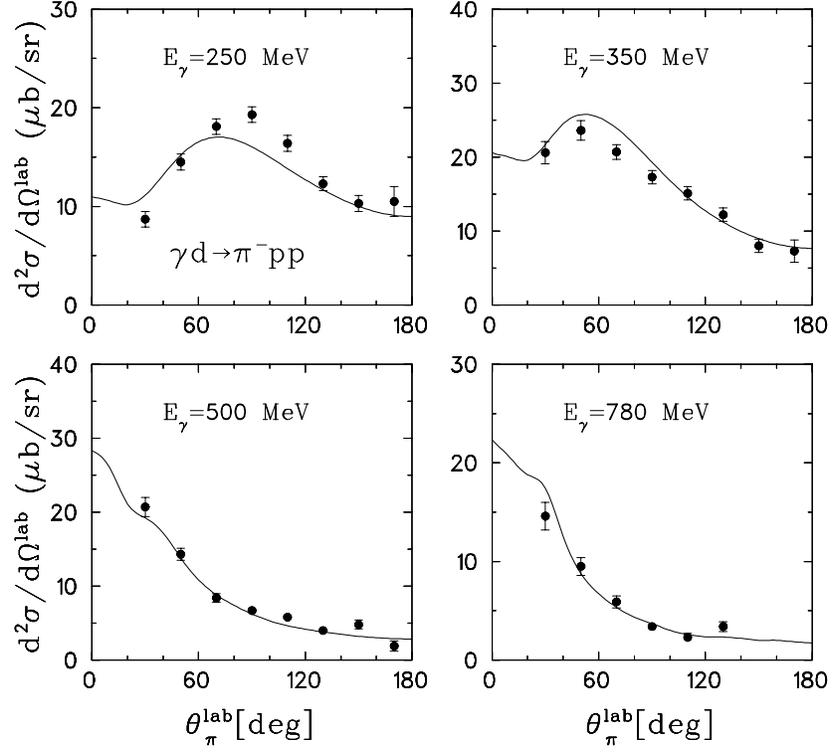}
\caption{Semi-exclusive differential cross section for
\protect$\pi^-$-photoproduction on the deuteron. 
Solid curves: IA +\protect$NN$-rescattering. Experimental data
from Benz et al.~\protect\cite{Be+73}.} 
\label{fig_piMi_diff_exp}
\end{figure}

\begin{figure}[htb]
\includegraphics[scale=.7]{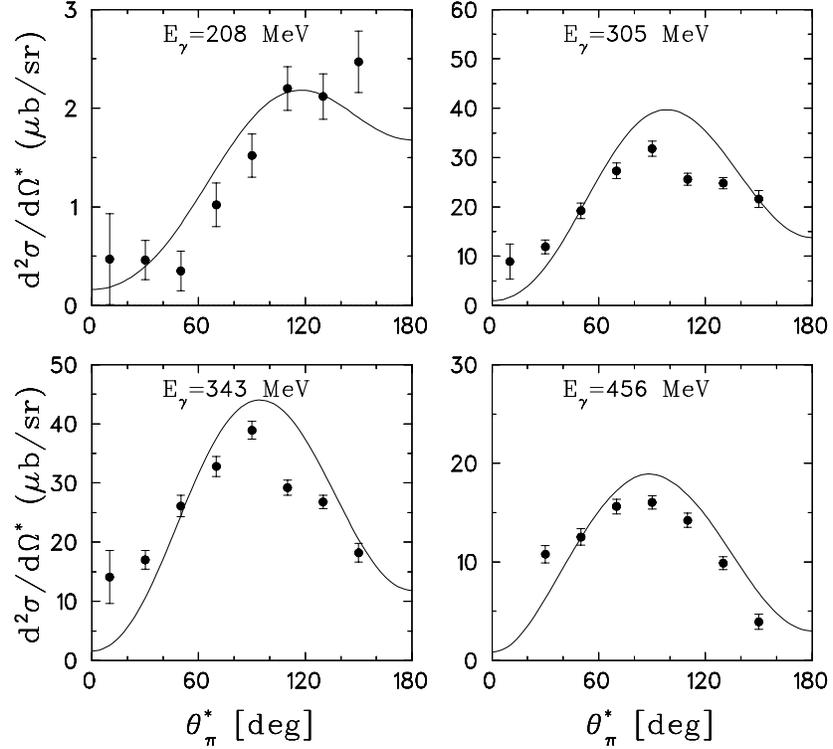}
\caption{Semi-exclusive differential cross section for
\protect$\pi^-$-photoproduction on the deuteron. Cross
section and the pion angle $\theta_\pi^*$ refer to the equivalent 
$\gamma N$ c.m.\ system of the corresponding reaction on the nucleon.
Solid curves: IA +\protect$NN$-rescattering. Experimental data
from Krusche et al.~\protect\cite{Kru99}.} 
\label{fig_pi0_diff_exp}
\end{figure}

\begin{figure}[htb]
\includegraphics[scale=.6]{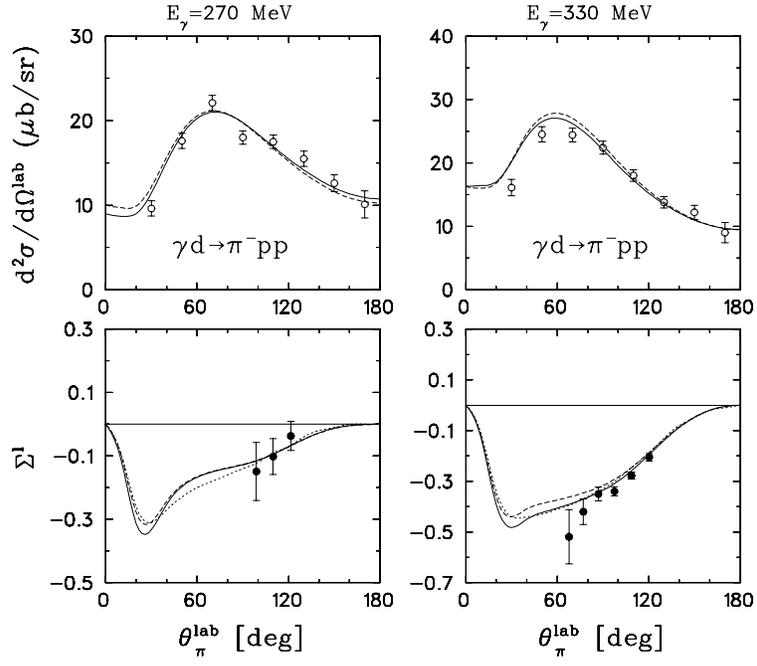}
\caption{Differential cross section and linear photon polarization
asymmetry for semi-exclusive \protect$\pi^-$ 
photoproduction on the deuteron. Notation of curves: dashed: IA;
dotted: IA from Lee and Sato~\protect\cite{Lee}; solid: IA +
\protect$NN$-rescattering. The data are for the differential cross
section from Benz et al~\protect\cite{Be+73} and for the linear photon
asymmetry from the LEGS-collaboration
(LEGS-exp.L3b)~\protect\cite{Sandorfi}.} 
\label{fig_pimi_sig}
\end{figure}

\begin{figure}[htb]
\includegraphics[scale=.9]{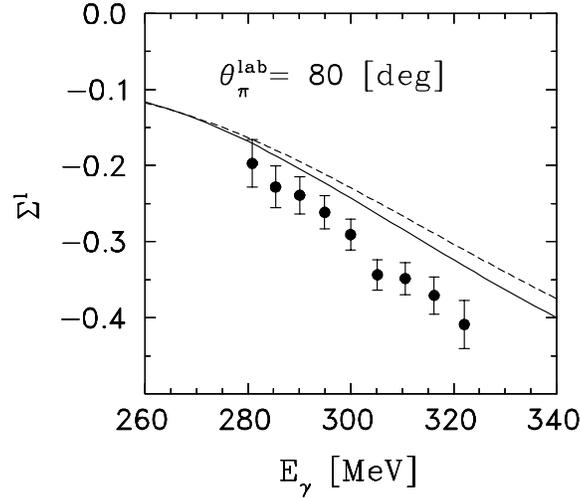}
\caption{Linear photon polarization asymmetry for semi-exclusive
\protect$\pi^-$-photoproduction on the deuteron at constant pion emission
angle \protect$\theta_\pi=80^\circ$ as function of photon energy. Notation of
curves: dashed: IA; solid: IA + \protect$NN$--rescattering. The
data are from the LEGS-collaboration 
(LEGS-exp.L3b)~\protect\cite{Sandorfi}.
} 
\label{fig_pimi_sig_e}
\end{figure}

\begin{figure}[htb]
\includegraphics[scale=.7]{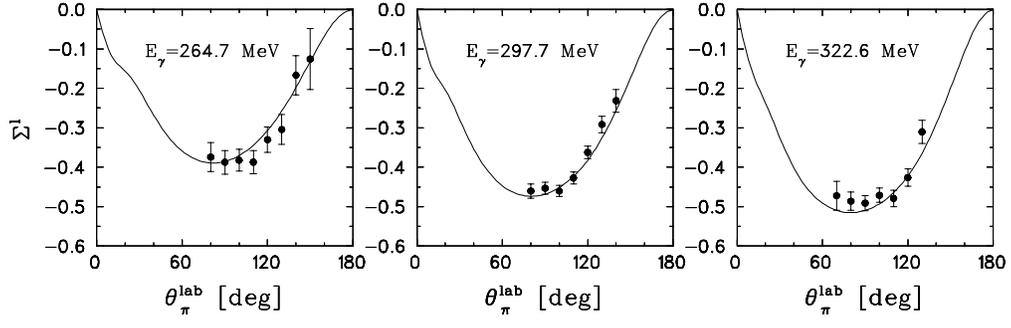}
\caption{Linear photon polarization
asymmetry for semi-exclusive \protect$\pi^0$-photoproduction on the
deuteron. 
Solid curves: IA + \protect$NN$-rescattering. Experimental data from the
LEGS-collaboration (LEGS-exp.L3b)~\protect\cite{Sandorfi}.} 
\label{fig_pi0_sig}
\end{figure}

\begin{figure}[htb]
\includegraphics[scale=.7]{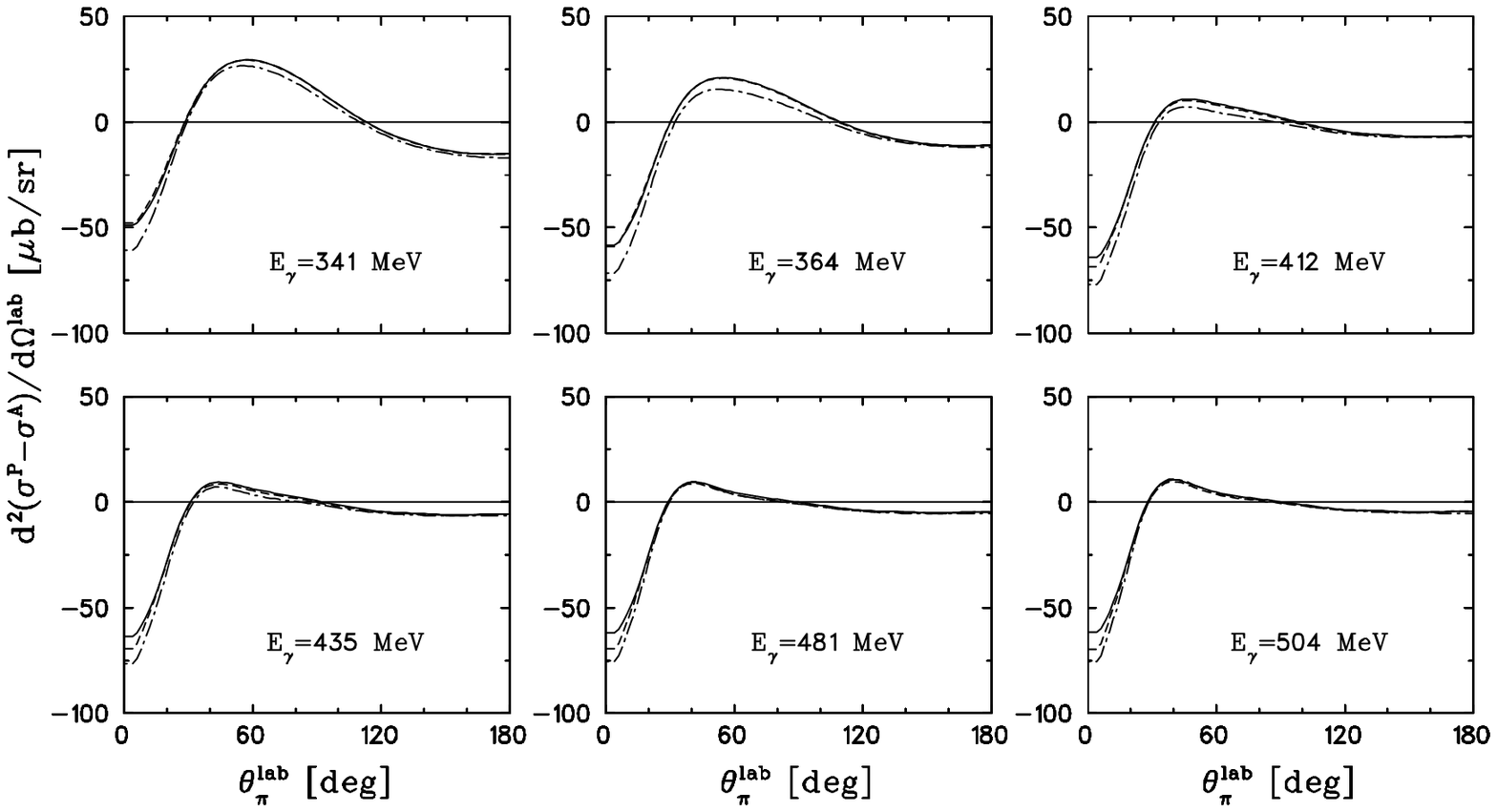}
\caption{Circular photon polarization asymmetry for semi-exclusive
\protect$\pi^-$-photoproduction on the deuteron. Notation of curves: 
dashed: IA; solid: IA + \protect$NN$-rescattering; dash-dot: 
corresponding elementary cross section.} 
\label{fig_piMi_GDH}
\end{figure}


\begin{thebibliography}{99}

\bibitem{ChL51}
        G.F. Chew and H.W. Lewis, Phys.\ Rev.\ {\bf 84}, 779 (1951).

\bibitem{LaF52}
        M. Lax and H. Feshbach, Phys.\ Rev.\ {\bf 88}, 509 (1952).

\bibitem{BlL77}
        I. Blomqvist and J.M. Laget, 
        Nucl.\ Phys.\ A {\bf 280}, 405 (1977).

\bibitem{Lag78}
        J.M. Laget, Nucl.\ Phys.\ A {\bf 296}, 388 (1978).

\bibitem{Lag81}
        J.M. Laget, Phys.\ Rep.\ {\bf 69}, 1 (1981).

\bibitem{Be+73}
        P. Benz {\it et al.}, Nucl.\ Phys.\ B {\bf 65}, 158 (1973).

\bibitem{LeP96}
        M.I. Levchuk, V.A. Petrun'kin, and M. Schumacher, 
        Z.\ Phys.\ A {\bf 355}, 317 (1996).   

\bibitem{LeS00}
        M.I. Levchuk, M. Schumacher, and F. Wissmann, nucl-th/0011041.

\bibitem{Said} 
        R.A. Arndt {\it et al.}, SAID: http://gwdac.phys.gwu.edu/.

\bibitem{Maid} 
        D. Drechsel, O. Hanstein, S.S. Kamalov, and L. Tiator, 
        MAID: http://www.kph.uni-mainz.de/de/MAID/maid2000/.

\bibitem{MaH87}
        R. Machleidt, K. Holinde, and Ch. Elster,
        Phys.\ Rep.\ {\bf 149}, 1 (1987); 
        R. Machleidt, 
        Adv.\ Nucl.\ Phys.\ {\bf 19}, 189 (1989).

\bibitem{Le+00}
        M.I. Levchuk, M. Schumacher, and F. Wissmann, 
	Nucl. Phys. A {\bf 675}, 621 (2000).

\bibitem{LoS00}
        A. Loginov, A.Sidorov, and V. Stibunov, 
	Phys. Atom. Nucl. {\bf 63},391 (2000) (Yad. Fiz. {\bf 63}, 459 (2000)).

\bibitem{DaA03a}
	E.M. Darwish, H. Arenh\"ovel, and M. Schwamb, Eur. Phys. J. A 
	{\bf 16}, 111 (2003).

\bibitem{DaA03b}
	E.M. Darwish, H. Arenh\"ovel, and M. Schwamb, Eur. Phys. J. A 
	{\bf 17}, 513 (2003).

\bibitem{GDH}
        Proc. Second Int. Symposium on the GDH sum rule and the Spin 
	Structure of the Nucleon,
	Genova 2002, eds. M. Anghinolfi, M. Battaglieri, and R. de Vita 
	(World Scientific, Singapore 2003);
        Proc. Third Int. Symposium on the GDH Sum Rule and its Extensions,
	Norfolk, Virginia, 2004, eds. S. Kuhn and J.-P. Chen,
	(World Scientific, Singapore 2005).

\bibitem{ScA96}
        R. Schmidt, H. Arenh\"ovel, and P. Wilhelm, 
	Z.\ Phys.\ A {\bf 355}, 421 (1996). 

\bibitem{Dar04a}
	E.M. Darwish, J. Phys. G {\bf 31}, 105 (2005).

\bibitem{Dar05a}
	E.M. Darwish, Nucl. Phys. A {\bf 735}, 200 (2005).

\bibitem{Dar05b}
	E.M. Darwish, Nucl. Phys. A {\bf 748}, 596 (2005).

\bibitem{Dar05c}
	E.M. Darwish, nucl-th/0504031. 

\bibitem{DaS05}
	E.M. Darwish and A. Salam, nucl-th/0505002.

\bibitem{ArF05}
	H. Arenh\"ovel and A. Fix, to be published.

\bibitem{ArF04}
	H. Arenh\"ovel, A. Fix, and M. Schwamb, Phys.\ Rev.\ Lett. 
	{\bf 93}, 202301 (2004).

\bibitem{FiA97}
	A. Fix and H. Arenh\"ovel, Z. Phys. A {\bf 359}, 427 (1997).

\bibitem{HaP85} 
        J. Haidenbauer and W. Plessas, 
        Phys.\ Rev.\ C {\bf 30}, 1822 (1984); {\bf 32}, 1424 (1985).

\bibitem{NoB90} 
        S. Nozawa, B. Blankleider, and T.-S.H. Lee,
        Nucl.\ Phys.\ A {\bf 513}, 459 (1990).

\bibitem{ChG57} 
	G.F. Chew, M.L. Goldberger, F.E. Low, and
  	Y. Nambu, Phys. Rev. {\bf 106}, 359 (1957). 

\bibitem{SaA04}
	A. Salam and H. Arenh\"ovel, Phys.\ Rev.\ C {\bf 70}, 044008 (2004).

\bibitem{Nob67}
	J. Noble, Phys. Lett. B {\bf 67}, 39 (1977).

\bibitem{Sio01}
        U. Siodlaczek {\it et al.}, Eur. Phys. J. A {\bf 10},365 (2001).

\bibitem{As+90}
        M. Asai {\it et al.}, 
        Phys.\ Rev.\ C {\bf 42}, 837 (1990).

\bibitem{Kru99}
        B. Krusche {\it et al.}, 
        Eur.\ Phys.\ J.\ A {\bf 6}, 309 (1999).    

\bibitem{ReA05}
	C. Reiss, H. Arenh\"ovel, and M. Schwamb, nucl-th/0505030.

\bibitem{BoC79}
        E. C. Booth, B. Chasan, J. Comuzzi, and P. Bosted, 
	Phys. Rev. C {\bf 20}, 1217 (1979).

\bibitem{LeB05}
        V. Lensky, V. Baru, J. Haidenbauer, C. Hanhart, A.E. Kudryavtsev,
	and U.-G. Meissner, nucl-th/0505039.

\bibitem{KoA87}
        G. K\"obschall, B. Alberti, H. Jansen, K. Rohrich, C. Schmitt, 
	V.H. Walther, K. Weinand, M. Kobayashi, and H. Arenh\"ovel, 
	Nucl. Phys. A {\bf 466}, 612 (1987).

\bibitem{LeC04}
        L. Levchuk, L. Canton, and A. Shebeko, 
	Eur. Phys. J. A {\bf 21}, 29 (2004), [nucl-th/0311004].

\bibitem{Sandorfi}
	A. Sandorfi, private communication.

\bibitem{Lee}
	T.-S. H. Lee, private communication.

\bibitem{Pedroni}
	P. Pedroni, private communication.

\bibitem{Bacci}
        C. Bacci {\it et al.}, Phys. Lett. B {\bf 39}, 559 (1972).

\bibitem{Hemmi}
        Y. Hemmi {\it et al.}, Nucl. Phys. B {\bf 55}, 333 (1975).

\bibitem{ArL05}
  H. Arenh\"ovel, W. Leidemann, and E.L. Tomusiak, Eur. Phys. J. A 
	{\bf 23}, 147 (2005). 

\end{thebibliography}
\end{document}